\documentclass{aastex61}
\usepackage{amsmath}
\usepackage{todonotes}
\usepackage{lineno}

\defcitealias{marble10}{M10}
\usepackage{color}
\submitjournal{ApJ}

\shorttitle{H{\sc ii} regions in N44}
\shortauthors{Barman et al.}

\begin{document}

\title{A study of photoionized gas in two H{\sc ii} regions of the N44 complex in the LMC using MUSE observations}

\correspondingauthor{Naslim Neelamkodan} 
\email{naslim.n@uaeu.ac.ae}

\author{Susmita Barman}
\affiliation{Department of Physics, College of Science, United Arab Emirates University (UAEU), Al-Ain, UAE, 15551.}
\affiliation{School of Physics, University of Hyderabad, Prof. C. R. Rao Road, Gachibowli, Telangana, Hyderabad, 500046, India}
\author[0000-0001-8901-7287]{Naslim Neelamkodan}
\affiliation{Department of Physics, United Arab Emirates University, Al-Ain, UAE, 15551.}
\author[0000-0003-3229-2899]{Suzanne C. Madden}
\affiliation{Laboratoire AIM, CEA/DSM - CEA Saclay, 91191 Gif-sur-Yvette, France}
\author{Marta Sewilo}
\affiliation{CRESST II and Exoplanets and Stellar Astrophysics Laboratory, NASA Goddard Space Flight Center, Greenbelt, MD 20771, USA}
\affiliation{Department of Astronomy, University of Maryland, College Park, MD 20742, USA}
\author[0000-0003-2743-8240]{Francisca Kemper}
\affiliation{European Southern Observatory, Karl-Schwarzschild-Str. 2, 85748, Garching b. München, Germany}
\affiliation{Institute of Astronomy and Astrophysics, Academia Sinica, 11F of Astronomy-Mathematics Building, AS/NTU, No.1, Sec. 4, Roosevelt Rd, Taipei 10617, Taiwan}
\author[0000-0002-2062-1600]{Kazuki Tokuda}
\affiliation{Department of Physical Science, Graduate School of Science, Osaka Prefecture University, 1-1 Gakuen-cho, Sakai, Osaka 599-8531, Japan}
\affiliation{National Astronomical Observatory of Japan, National Institutes of Natural Science, 2-21-1 Osawa, Mitaka, Tokyo 181-8588, Japan}
\author{Soma Sanyal}
\affiliation{School of Physics, University of Hyderabad, Prof. C. R. Rao Road, Gachibowli, Telangana, Hyderabad, 500046, India}
\author{Toshikazu Onishi}
\affiliation{Department of Physical Science, Graduate School of Science, Osaka Prefecture University, 1-1 Gakuen-cho, Sakai, Osaka 599-8531, Japan}

\begin{abstract}

We use the optical integral field observations with Multi-Unit Spectroscopic Explorer (MUSE) on the Very Large Telescope, together with CLOUDY photoionization models to study ionization structure and physical conditions of two luminous H\,{\sc ii} regions in N44 star-forming complex of the Large Magellanic Cloud. The spectral maps of various emission lines reveal a stratified ionization geometry in N44\,D1. The spatial distribution of [O\,{\sc i}]\,6300\AA\,emission in N44\,D1 indicates a partially covered ionization front at the outer boundary of the H\,{\sc ii} region. These observations reveal that N44\,D1 is a Blister H\,{\sc ii} region. The [O\,{\sc i}]\,6300\AA\, emission in N44\,C does not provide a well-defined ionization front at the boundary, while patches of [S\,{\sc ii}]\,6717\AA\, and [O\,{\sc i}]\,6300\AA\, emission bars are found in the interior. The results of spatially resolved MUSE spectra are tested with the photoionization models for the first time in these H\,{\sc ii} regions. A spherically symmetric ionization-bounded model with a partial covering factor, which is appropriate for a Blister H\,{\sc ii} region can well reproduce the observed geometry and most of the diagnostic line ratios in N44\,D1. Similarly, in N44\,C we apply a low density and optically thin model based on the observational signatures. Our modeling results show that the ionization structure and physical conditions of N44\,D1 are mainly determined by the radiation from an O5\,V star. However, local X-rays, possibly from supernovae or stellar wind, play a key role. In N44\,C, the main contribution is from three ionizing stars. 


\end{abstract}
\section{Introduction}
\label{sec:introduction}
  


Massive stars are the significant sources of ultraviolet (UV) radiation in galaxies with energies high enough ($>$\,13.6\,eV) to ionize the neutral gas in the interstellar medium (ISM). A part of this high-energy radiation is absorbed by the neutral gas and heats the surrounding medium creating the ionized H\,{\sc ii} regions. A large fraction of this ionizing radiation escapes into the diffuse medium, penetrating the molecular gas, if some part of the H\,{\sc ii} region is optically thin. This creates an ionization zone (H$^+$), ionization front (H$^0$), and a photodissociation region (PDR). The ionization front is at the outer boundary of an H\,{\sc ii} region that lies inside the PDR. There have been several studies of Galactic and extragalactic H\,{\sc ii} regions and PDRs, e.g., Orion Nebula \citep{pogge92, garcia07}, 30\,Doradus \citep{pellegrini10_30dor}, NGC\,364 \citep{peimbert00, Relano_2002}, dense H\,{\sc ii} regions in IC\,10 \citep{polles19} and NGC\,595 \citep{relano10}. The impact of ionizing radiation on the surrounding medium and the physical properties of H\,{\sc ii} regions are normally obtained by strong emission lines in the optical spectrum, which is mainly populated by hydrogen recombination lines and forbidden lines of other common elements. These gas emission lines are sensitive to physical conditions such as density and temperature; hence their relative intensities can probe the physical mechanism involved in the ionization processes.  



The Large Magellanic Cloud (LMC) is an ideal laboratory to study the properties of H\,{\sc ii} regions and the massive star feedback in a low-metallicity galaxy due to its sub-solar metallicity ( Z\,=\,0.5Z$_\odot$, \citealt{westerlund97}); face-on viewing angle \citep{van01}; reduced extinction along the line of sight; a distance of 50\,kpc \citep{pietrzynki19} allowing the spatially resolved observations of the ISM structures on sub-parsec scales. \citet{Naslim15} have reported on ten PDRs in the LMC using H$_2$ pure rotational transition emission obtained with {\it Spitzer}. These regions include intense H\,{\sc ii} regions, diffuse ISM clouds, and dense molecular clouds. We study the individual clouds in detail using various observations to investigate the high-mass star formation (see \citealt{naslim18, nayana20}) and its impact on the ISM. To explore the impact of ionizing radiation from massive stars on the surrounding medium, we revisit two H\,{\sc ii} regions in a well-studied star-forming complex of the LMC, N44. We examine the physical conditions and ionization structure of N44D and N44C by comparing the observations  with the predictions of the photoionization model, CLOUDY \citep{ferland2017}. 

\begin{figure*}
    \centering
    \includegraphics[width= 150 mm, height= 120 mm]{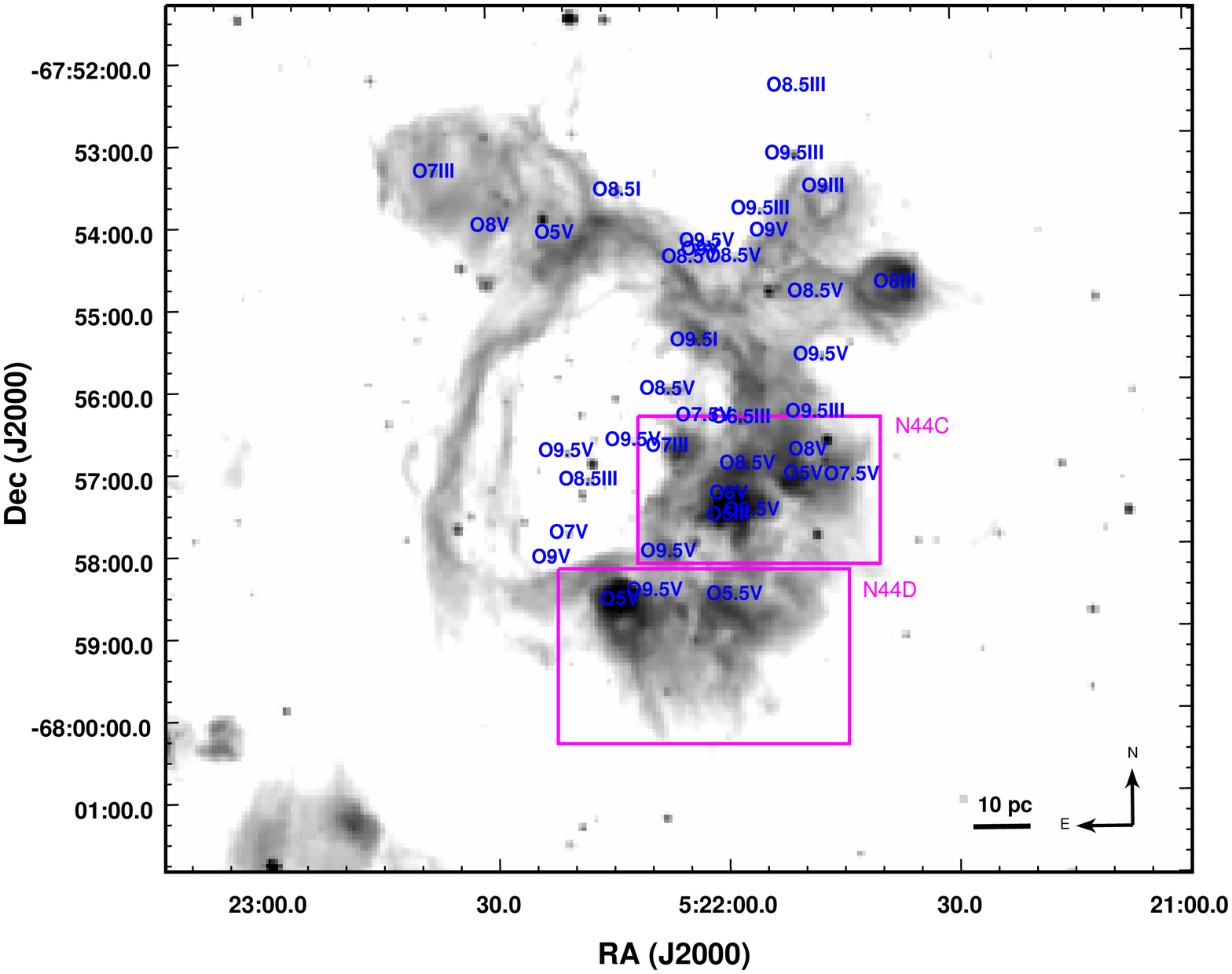}
    \caption{The H$\alpha$ map of whole N44 superbubble obtained from Magellanic Cloud Emission Line Survey (MCELS; \citealt{smith98}). Spectral type of hot stars \citep{McLeod19} are labeled on the H$\alpha$ emission map and the magenta coloured boxes indicate the N44\,D and N44\,C H\,{\sc ii} regions. }
    \label{N44_OB}
\end{figure*}

 The N44 superbubble is one of the brightest star-forming regions in the LMC, which can be clearly traced by its compact H\,{\sc ii} regions along the main shell rim in an H$\alpha$ map (Fig.\,\ref{N44_OB}). The region is powered by nearly 35$-$38 hot stars \citep{Oye_Massey1995, McLeod19} of stellar associations LH47, LH48 and LH49 \citep{lucke70}.  \citet{McLeod19} report the analysis of radiative and mechanical feedback from massive stars in the H\,{\sc ii} regions of N44 (N44\,A, N44\,B, N44\,C and N44\,D) using the optical integral field data from Multi-Unit Spectroscopic Explorer (MUSE). They used He\,{\sc ii}\,5412\AA\, line to identify the feedback driving massive stars, and estimated the spectral types and luminosity classes of these stars for determining the stellar radiative output. Using the nebular emission line maps of H$\alpha$, H$\beta$, [S\,{\sc ii}] 6717, 6732\AA, [N\,{\sc ii}] 6584\AA, and [O\,{\sc iii}] 5007\AA, they derived the electron density from [S\,{\sc ii}] 6717/6732 ratio assuming a nebular electron temperature of 10\,000\,K. They also derived the degree of ionization using [O\,{\sc ii}]/[O\,{\sc iii}] ratio, kinematics, and the oxygen abundances. In addition, they explored the role of different stellar feedback mechanisms by estimating various pressures and found that the H\,{\sc ii} region expansion is mainly driven by stellar radiation pressure and ionized gas. 
 
  In this paper, we further explore the rich nebular emission lines in MUSE data set \citep{McLeod19} of N44 for a detailed understanding of their spatial distributions, ionization structures and physical conditions. We compare these results with the photoionization models to interpret the observations of this star-forming complex. The MUSE archival data of N44 provides many iconic emission lines, such as H$\alpha$, H$\beta$, [S\,{\sc iii}] 9069\AA, [S\,{\sc ii}] 6717, 6732\AA, [O\,{\sc iii}] 4959, 5007\AA, [O\,{\sc ii}] 7318, 7329\AA, [O\,{\sc i}] 6300\AA, and [N\,{\sc ii}] 6584\AA, that have only been partially utilized in \citet{McLeod19}. We exploit various emission line ratios to study the behavior of different ionization zones and compare the observed line ratios with the best-fit photoionization models to further understand the physical process. 
  
  The N44 main shell is surrounded by several H$\alpha$ bright regions (Fig.\,\ref{N44_OB}). The compact H\,{\sc ii} region on the southwest rim of the shell, N44\,D, is the most luminous one in N44 \citep{McLeod19}. We choose two brightest H\,{\sc ii} regions N44\,D and N44\,C for our study, which show higher degree of ionization than N44\,A and N44\,B, and nearly spherical ionization structures with different feedback characteristics based on studies by \citet{McLeod19}. N44\,D encloses three hot stars of spectral types O5\,V, O9.5\,V and O5.5\,V \citep{McLeod19}. The second brightest H\,{\sc ii} region, N44\,C, is adjacent to N44D on the western rim of the shell, and the region harbors three hot stars O5\,III, O9.5\,V and O5\,III \citep{McLeod19}. One reason for selecting N44\,D is that MUSE observations show an edge on ionization front, allowing a detailed study of ionization as a function of depth into the cloud. At first glance, N44\,C does not show such a well-defined ionization front at the H\,{\sc ii} region boundary, while it shows a higher photon escape fraction relative to N44\,D that reported by \citet{McLeod19}. These two nebulae show different ionization structures, hence physical process of two different types of H\,{\sc ii} regions can be compared. Moreover, the hot star content of N44 H\,{\sc ii} regions are extensively studied and high-spatial-resolution observations are readily available. This study allows us to directly apply the observed stellar parameters, gas densities, and emission line intensities to constrain the photoionization model without arbitrary assumption, and test their influence on geometry of the H\,{\sc ii} regions.
 
\section{Observations}
We used the MUSE archival data of N44\,C and N44\,D (program ID: 096.C-0137(A), PI: A. F. McLeod). MUSE is a large field-of-view (FOV) integral field unit (IFU) panchromatic optical instrument on the European Southern Observatory's (ESO's) Very Large Telescope (VLT) in Paranal, Chile. This instrument provides high-spatial-resolution observations at a pixel scale of 0.2$^{\prime\prime}$ with a resolving power ranging from 1770 to 3590. The observations of N44\,C and N44\,D have been taken on 21 October 2015 and 25 February 2016, with the MUSE$\_$wfm-noao$\_$obs$\_$genericoffset observing template, in a wide-field observing mode covering a wavelength range 475$-$935\,nm. The reduced MUSE data is retrieved from the ESO science archive\footnote{http://archive.eso.org/cms.html}. The data were reduced using the MUSE-1.6.1 pipeline. The MUSE pipeline process automatically removes most of the instrumental signatures. The raw data were pre-processed, and bias subtraction, flat fielding, sky-subtraction, wavelength calibration and flux calibration were applied. These data were not taken with the Adaptive Optics System of MUSE and the seeing-limited angular resolutions 0.98$^{\prime\prime}$ and 1.30$^{\prime\prime}$ are achieved for N44D and N44C respectively. We note that, no point spread function (PSF) matching was applied for subsequent analysis and all the analysed regions are resolved regardless of the achieved seeing. Our analysis is based on integrated line flux maps, hence no PSF information is retained. 

\section{Emission line maps of N44\,D and N44\,C}
\begin{figure*}
    \hspace*{0mm}\includegraphics[width=180mm, height=45mm]{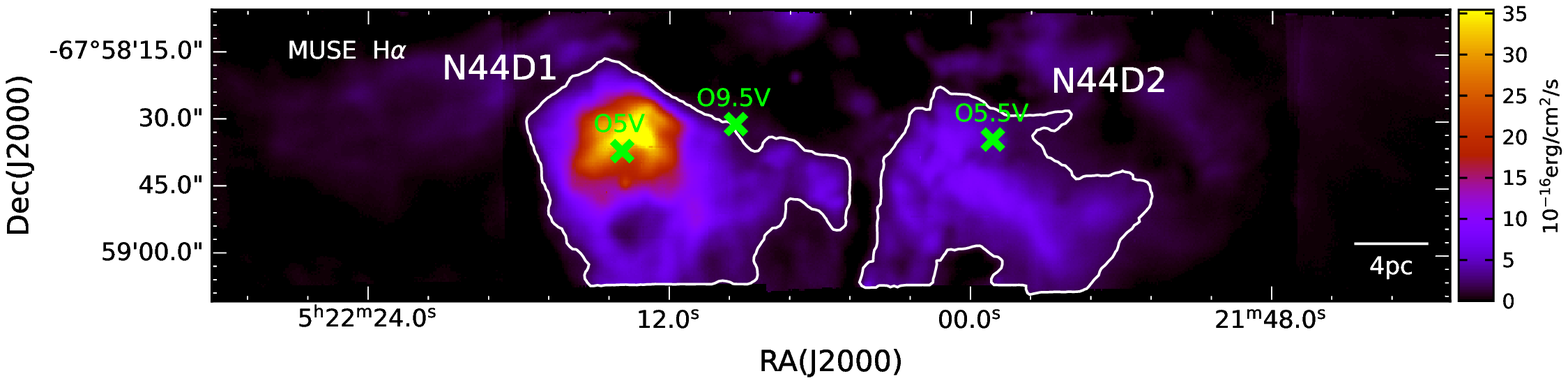}
    \hspace*{0mm}\includegraphics[width=180mm, height=45mm]{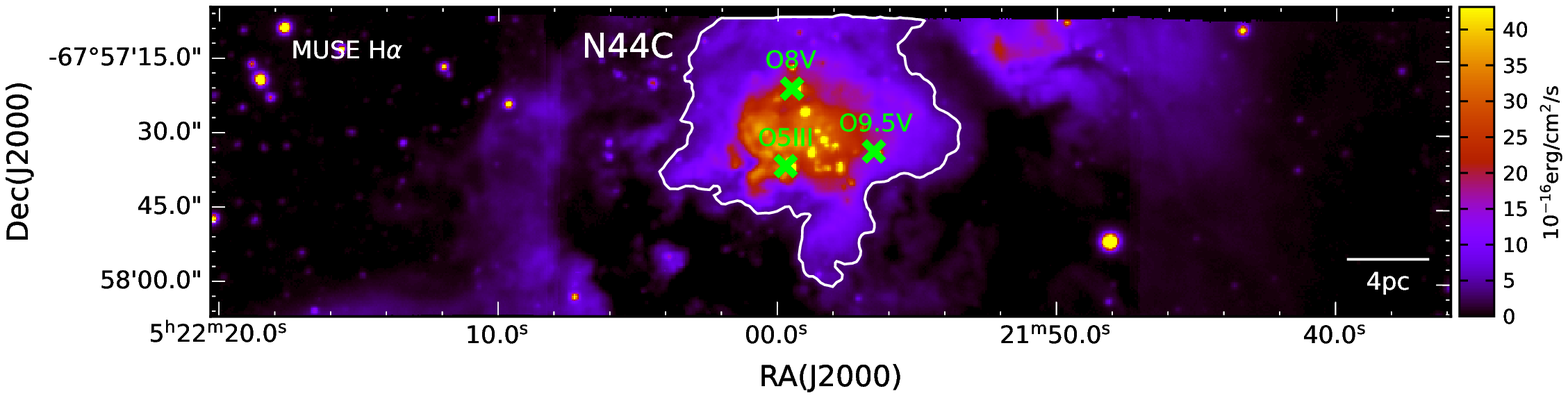}
    \caption{The MUSE H$\alpha$ maps of N44\,D (top) and N44\,C (bottom). The emission line fluxes are extracted from the polygon regions (white) of N44\,D1, N44\,D2 and N44\,C for the analysis of photoionized gas in this work.}
    \label{fig:Halpha}
\end{figure*}
Fig.\,\ref{fig:Halpha} shows the extinction corrected H$\alpha$ line flux maps of N44\,D and N44\,C obtained with MUSE. The ionized gas traced by the H$\alpha$ emission shows two H\,{\sc ii} regions in N44\,D, those we label as N44\,D1 and N44\,D2, and one in N44\,C. Even though the regions are nearly spherical, their boundaries cannot be directly specified in a circular aperture. Hence for determining the boundaries of these H$\alpha$ bright regions based on their surface brightness, we use the Python package ASTRODENDRO \citep{Rosolowsky08}. This algorithm identifies and characterizes the hierarchical structures in the emission line map as a structure tree, where each entity is represented as an isosurface. The local maxima represent the top level of the dendrogram and are identified from the emission line map with the flux $>$ 3$\sigma$. The isosurfaces (two-dimensional contours) that surround the local maxima are leaves, branches, and trunks. The trunks represent parent structures that enclose the branches connecting two leaves. Further description and methods of using ASTRODENDRO can be found in \citet{naslim18}. \citet{bruna20} have recently used ASTRODENRO to identify H$\alpha$ bright regions in MUSE maps of NGC\,7793 by applying a similar method. We define the boundary of H\,{\sc ii} regions in H$\alpha$ map that are identified as trunks with ASTRODENDRO. The lower contour levels of these trunks are taken as the boundaries of H\,{\sc ii} regions within the chosen observation field. These regions appear as polygons in Fig.\,\ref{fig:Halpha}. 
\begin{figure}
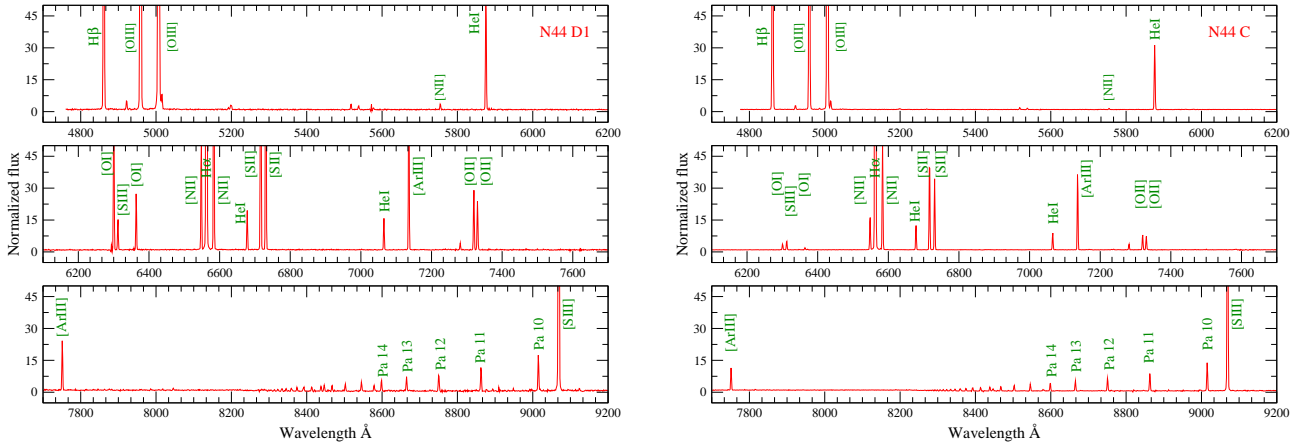

    \centering
\hspace{0mm}\includegraphics[scale=0.35]{N44D_lines_grace.eps}
\hspace{0.5 cm}
\hspace{0mm}\includegraphics[scale=0.35]{N44C_lines_grace.eps}
    \caption{The identified emission lines are labelled in the MUSE spectra extracted from a 1.0$^{\prime\prime}$ radius circular region close to the cross-cut indicated as red line in Fig.\,\ref{fig:spatialmap}. This is to show the rich emission lines available for analysis in the MUSE observation of N44\,D1 and N44\,C. The continua of the spectra are normalized to 1 and the peak of certain strong lines are cut out of the scale for the weaker lines to be visible in the plot properly.}
    \label{N44_lines}
\end{figure} 

In addition to H$\alpha$\,\,6562.8\AA\, emission, the MUSE spectra of N44\,D1 and N44\,C show emission due to H$\beta$\,4861\AA, [O\,{\sc iii}] 5007\AA, 4959\AA, [O\,{\sc ii}] 7318\AA, 7329\AA, [O\,{\sc i}] 6300\AA, [N\,{\sc ii}] 6584\AA, [S\,{\sc ii}] 6717\AA, 6732\AA, [S\,{\sc iii}] 9069\AA, [Ar\,{\sc iii}] 7135\AA, 7751\AA\, and many He\,{\sc i} and Paschen hydrogen lines ( Fig \ref{N44_lines}). Fig.\,\ref{fig:spatialmap} shows integrated line flux maps of H$\beta$, [S\,{\sc ii}]\,6717\AA, [O\,{\sc iii}]\,5007\AA, [O\,{\sc ii}]\,7318\AA, [O\,{\sc i}]\,6300\AA, and [N\,{\sc ii}]\,6584\AA\, in N44\,D1, N44\,D2, and N44\,C. H$\beta$ and [O\,{\sc iii}]\,5007\AA\, emission show similar spatial distribution as the H$\alpha$ emission in both N44\,D and N44\,C, whilst [S\,{\sc ii}], [O\,{\sc ii}], [N\,{\sc ii}] and [O\,{\sc i}] emission in N44\,D1 show a shell structure. The morphology of N44\,D2 is irregular with pillars or filamentary structures. 
The continuum subtraction is applied by creating continuum maps from the user defined line-free portions of the spectrum around each emission line. The line fluxes (erg s$^{-1}$ cm$^{-2}$) of N44\,D1, N44\,D2 and N44\,C regions are then extracted from the integrated line flux maps by applying aperture photometry within the specified regions as polygon structures obtained from ASTRODENDRO. Only the pixel values with S/N\,$>$\,10 are considered within all the polygons. These fluxes are then point source flux removed by subtracting the point source fluxes, which are also extracted within the user defined apertures. For uncertainties, we added in quadrature the error in aperture photometry, and an expected 20$\%$ calibration error in every line flux measurements. The error in photometry is the quadratically added uncertainty in measurements over all pixels within a region. 

To check the data reduction and flux calibration quality of data cube retrieved from the MUSE archive, we compared the H$\alpha$ line luminosities (erg s$^{-1}$) of N44\,C and N44\,D1 from the MUSE archival data to the H$\alpha$ luminosities presented in \citet{McLeod19}. In Table \ref{comparison} we show the comparison of H$\alpha$ luminosities obtained from the data presented in this work and \citet{McLeod19}. We note that the line luminosities obtained from the two data sets agree within the estimated uncertainties, hence we are confident to proceed with the analysis of pipeline reduced MUSE archival data. The observed line luminosities (erg s$^{-1}$) are given in Table \ref{lines}. We choose two bright regions, N44\,D1 and N44\,C, for further analysis with photoionization models. 

\begin{table*}
\centering
\caption{Comparison of H$\alpha$ luminosities obtained from the MUSE archival pipeline data (this work) with the data obtained from \citet{McLeod19}}.
\renewcommand{\arraystretch}{1.2}
\begin{tabular}{lcccccccc}
\hline
\multicolumn{2}{c}{N44\,D1}&
\multicolumn{2}{c}{N44\,C}&\\
$^{a}L_{\textrm{obs}}$(ergs$^{-1}$)&$^{b}L_{\textrm{obs}}$(erg s$^{-1}$)&$^{a}L_{\textrm{obs}}$(erg s$^{-1}$)&$^{b}L_{\textrm{obs}}$(erg s$^{-1}$)\\
$(\times$10$^{37}$)&($\times$10$^{37}$)&($\times$10$^{37}$)& ($\times$10$^{37}$) \\
\hline 
1.47$\pm$0.16&1.47&1.39$\pm$0.15&1.51 \\
\hline
\label{comparison}
\end{tabular}
\parbox{100mm}{
a: MUSE archival data used in this work.
b: \citet{McLeod19}.
}
\end{table*}


\section{Emission line ratios: [S\,{\scriptsize II}]/H$\alpha$, [N\,{\scriptsize II}]/H$\alpha$, [O\,{\scriptsize III}]/H$\alpha$ and [O\,{\scriptsize III}]/H$\beta$}

\begin{figure*}
 \centering
\includegraphics[scale=0.35]{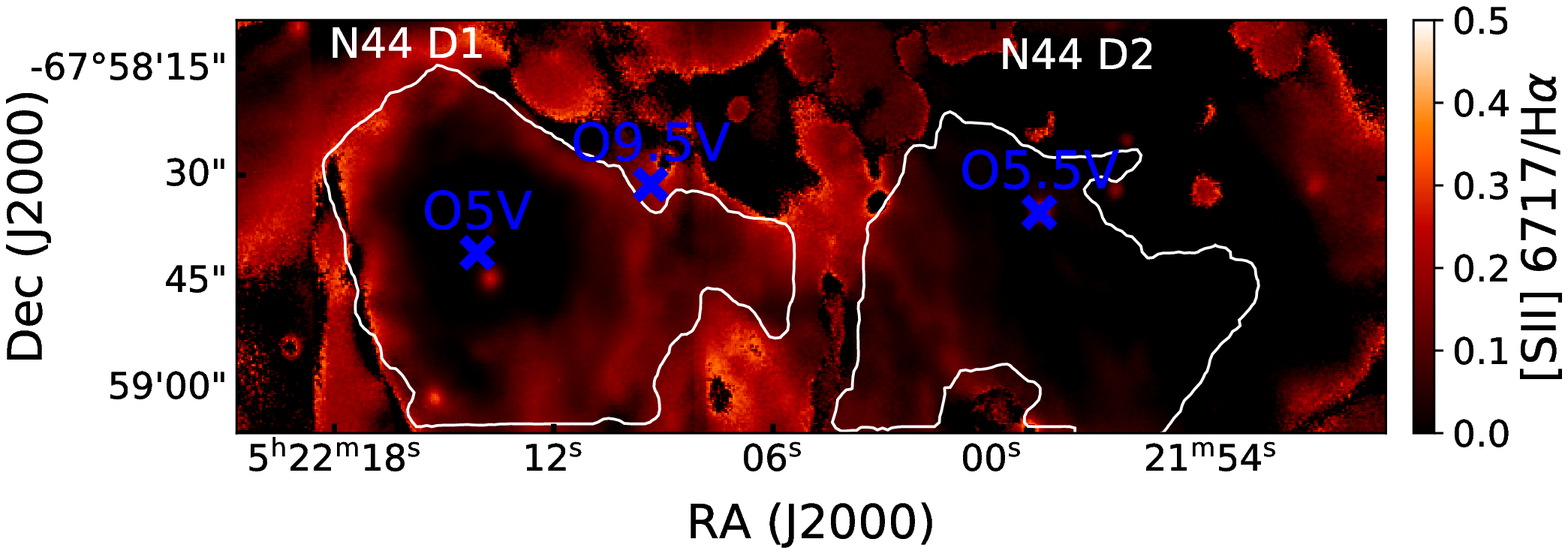}
\includegraphics[scale=0.35]{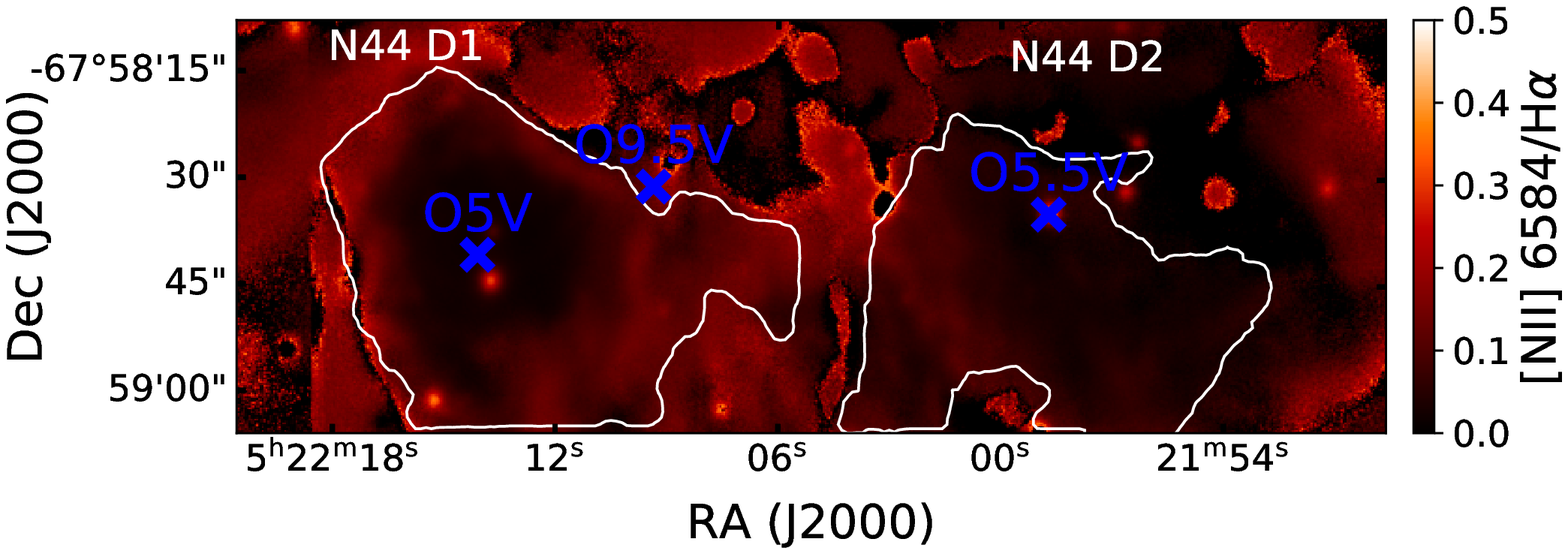}\\
\includegraphics[scale=0.35]{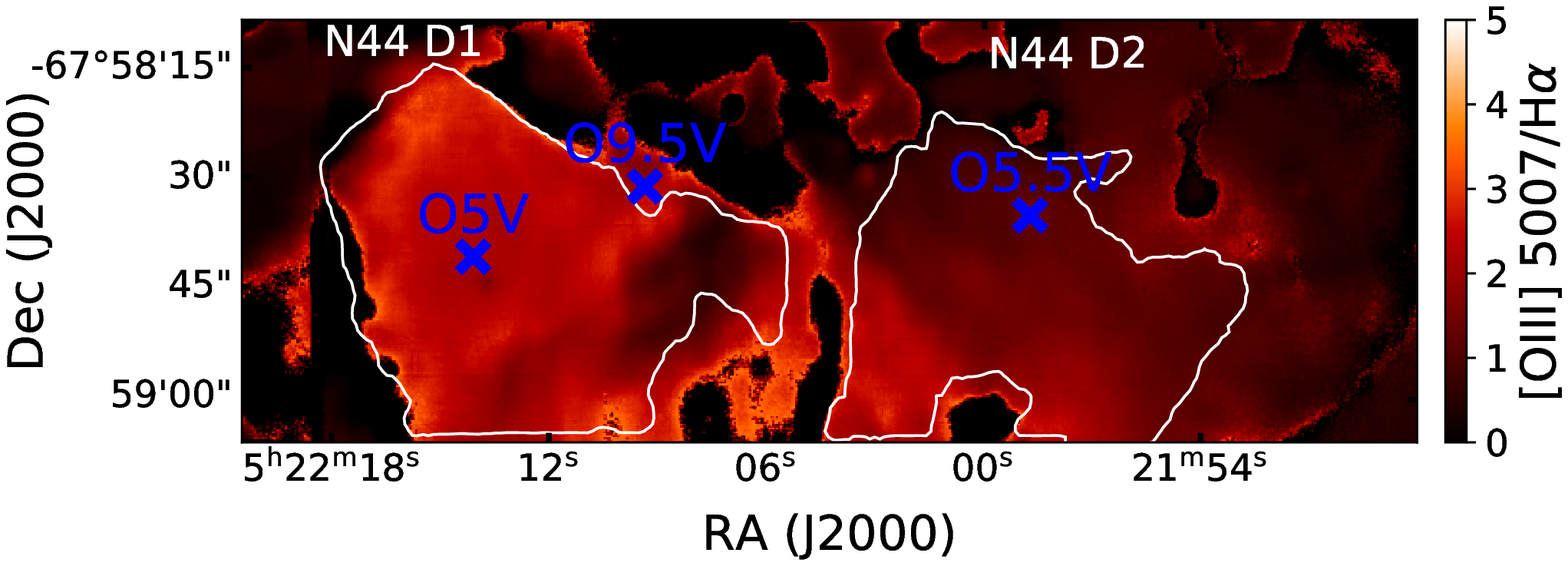}
\includegraphics[scale=0.35]{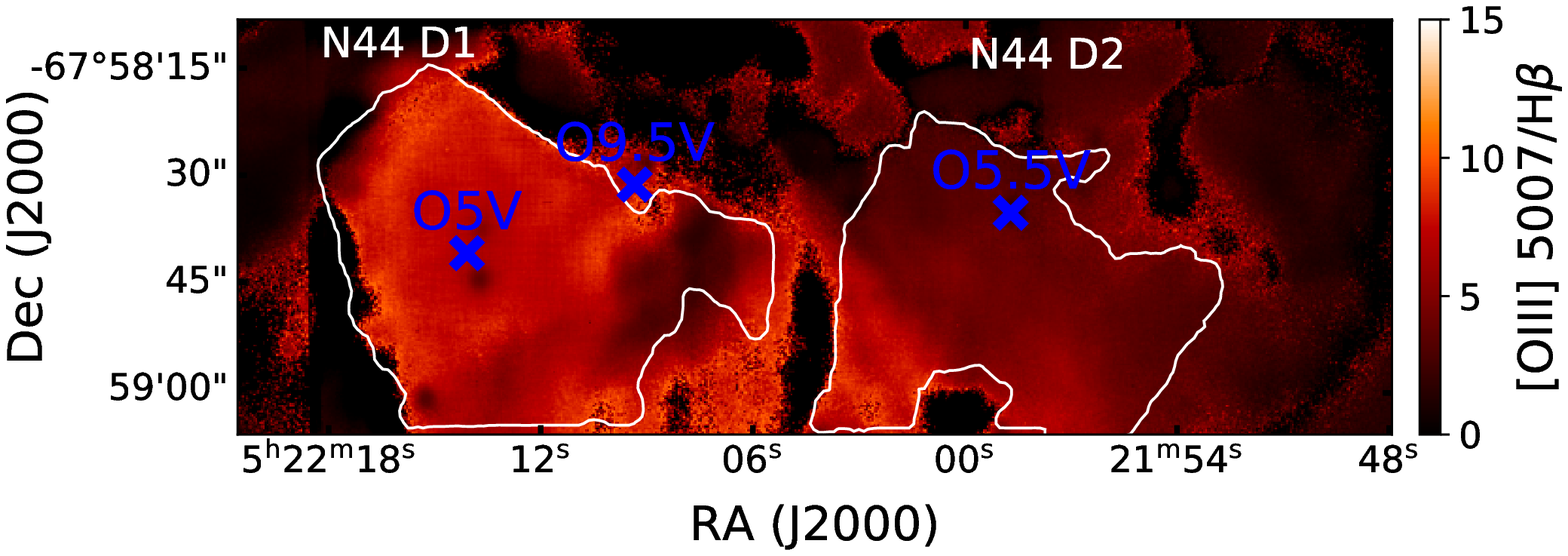} 
    \caption{
    [S\,{\sc ii}]6717/H$\alpha$, [N\,{\sc ii}]6584/H$\alpha$, [O\,{\sc iii}]\,5007/H$\alpha$ and [O\,{\sc iii}]\,5007/H$\beta$ ratio maps of N44\,D1 and N44\,D2 regions. The white polygons represent the regions taken for analysis in this work and blue cross labels are the locations of hot stars.}
    \label{lineratioN44D}
\end{figure*}

\begin{figure*}
    \centering
    
\includegraphics[scale=0.35]{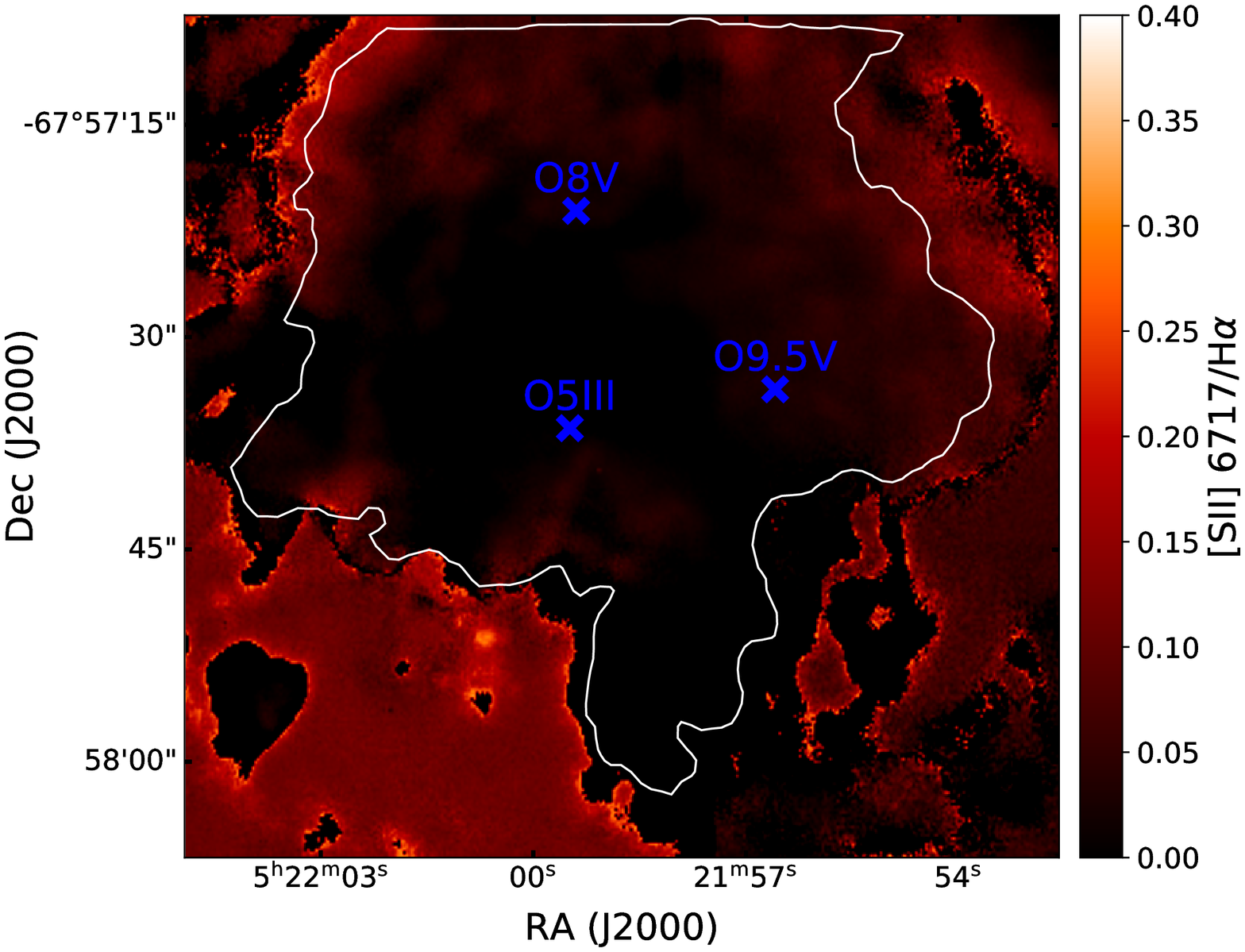}
\includegraphics[scale=0.35]{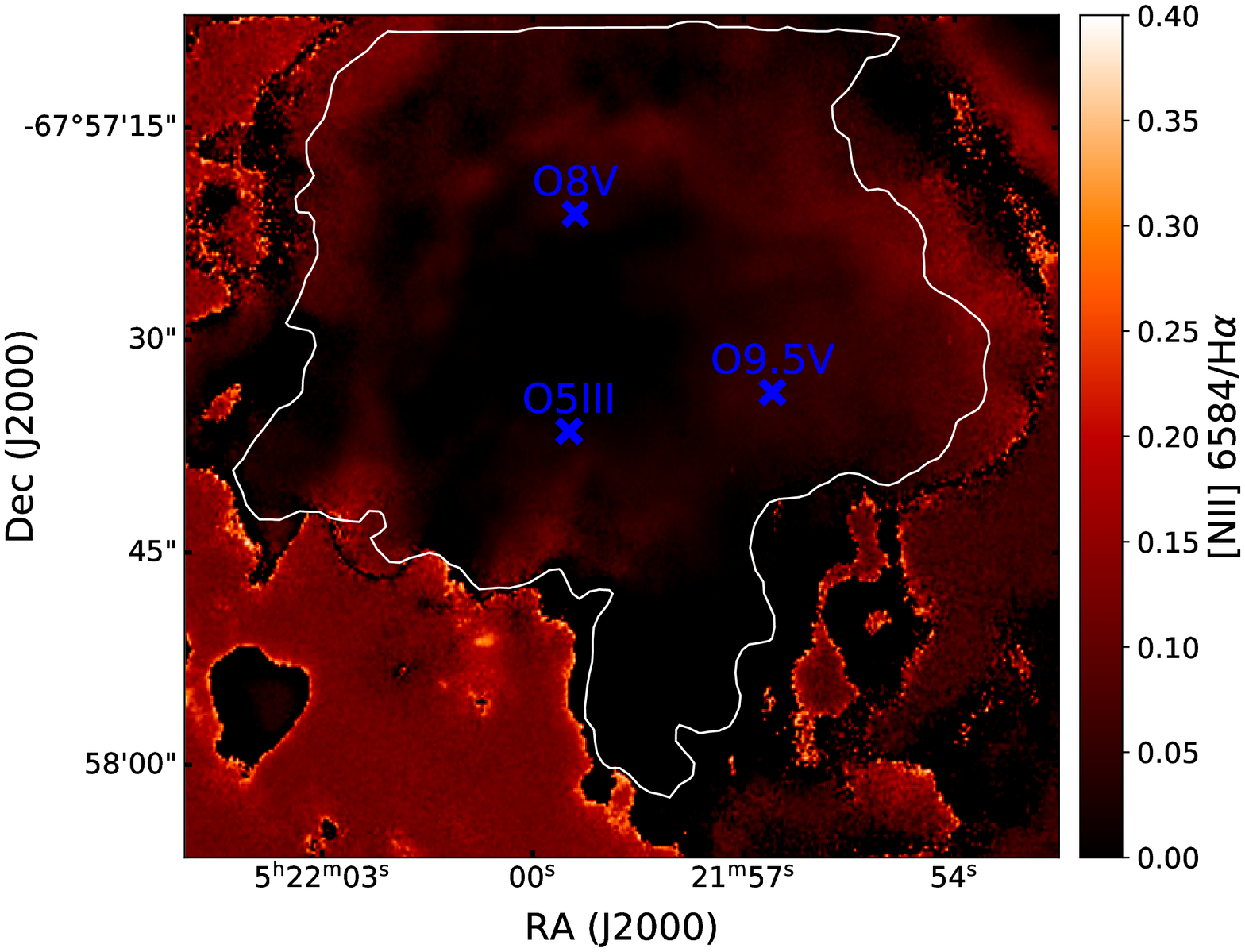}\\
\includegraphics[scale=0.35]{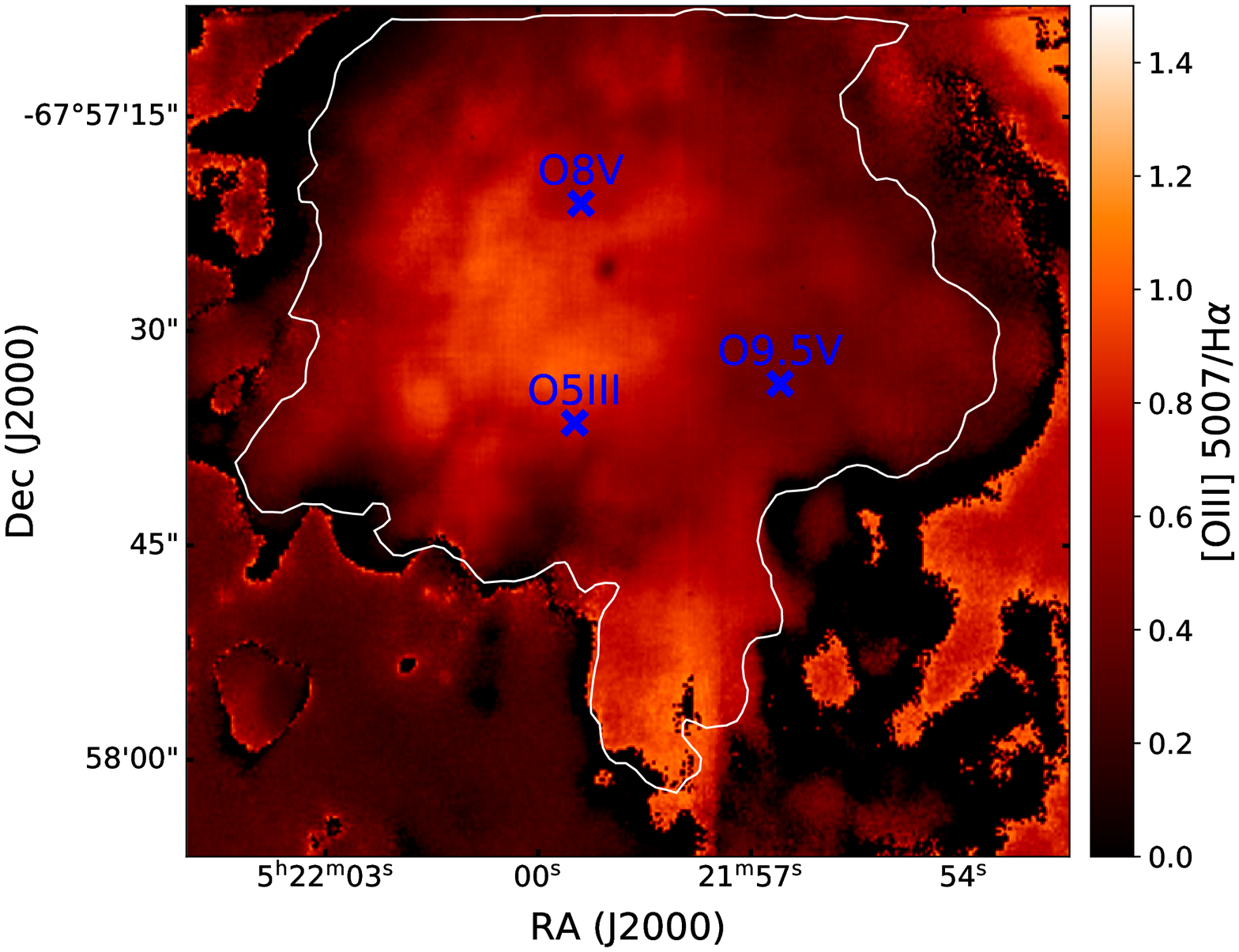}
\includegraphics[scale=0.35]{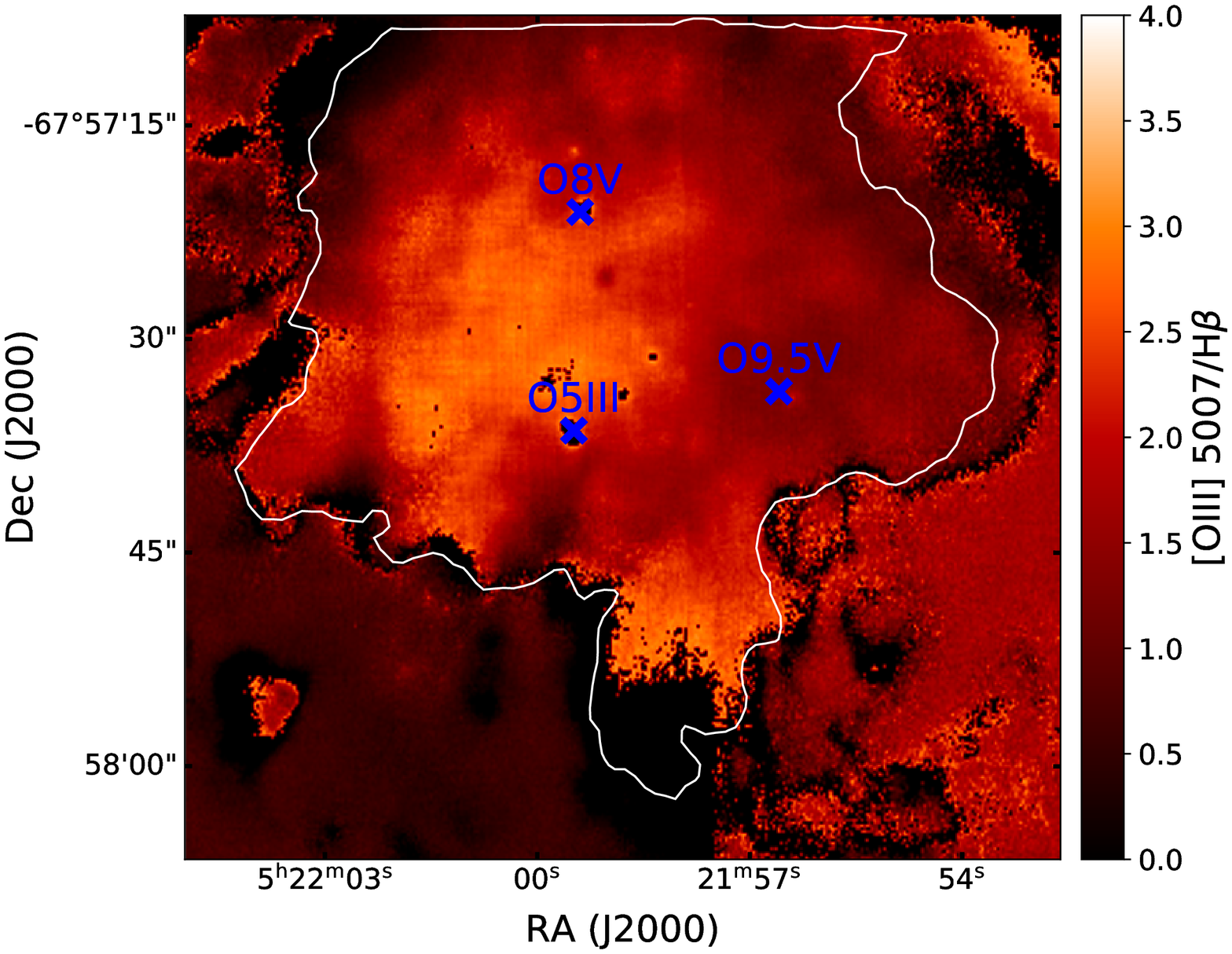}                                                                                                         

    \caption{
    [S\,{\sc ii}]\,6717/H$\alpha$, [N\,{\sc ii}]\,6584/H$\alpha$, [O\,{\sc iii}]\,5007/H$\alpha$ and [O\,{\sc iii}]\,5007/H$\beta$ ratio maps of N44\,C region. The white polygons represent the regions taken for analysis in this work and blue cross labels are the locations of hot stars.}
    
    \label{lineratioN44C}
\end{figure*}

We present the extinction corrected [S\,{\sc ii}]\,6717/H$\alpha$, [N\,{\sc ii}]\,6584/H$\alpha$, [O\,{\sc iii}]\,5007/H$\alpha$, and [O\,{\sc iii}]\,5007/H$\beta$ ratio maps of N44\,D and N44\,C in Figs.\,\ref{lineratioN44D} and \ref{lineratioN44C}, respectively. These ratios allow us to study the ionization structure of the region. [S\,{\sc ii}]\,6717/H$\alpha$ and [N\,{\sc ii}]\,6584/H$\alpha$ are lower at the central regions closer to the ionizing stars implying a higher ionization zone, while at the periphery the values of these ratios are higher indicating low ionization zone. We find that both [S\,{\sc ii}]\,6717/H$\alpha$ and [N\,{\sc ii}]\,6584/H$\alpha$ maps of N44\,D1 and N44\,C show a shell morphology. The central part of the N44\,D1 has a lower [S\,{\sc ii}]\,6717/H$\alpha$ ratio ($\sim$\,0.02), and at the periphery, its value is $\sim$\,0.3. N44\,C shows a [S\,{\sc ii}]\,6717/H$\alpha$ value of $\sim$\,0.04 at the center and a value of $\sim$\,0.20 at the periphery. Similarly, the value of [N\,{\sc ii}]\,6584/H$\alpha$ ratio ranges from 0.03 to 0.20 in N44\,D1, and 0.04 to 0.15 in N44\,C. The model calculations by \citet{Allen08} have shown that [S\,{\sc ii}]/H$\alpha$ and [N\,{\sc ii}]/H$\alpha$ ratios greater than 0.39 and 0.79 would be a result of strong contributions from shocks. The [S\,{\sc ii}]/H$\alpha$ and [N\,{\sc ii}]/H$\alpha$ ratios of both N44\,D1 and N44\,C are well below the values 0.39 and 0.79, respectively, indicating a substantial contribution from photoionization. However, in the regions outside the boundary of these H\,{\sc ii} regions, we find the enhanced [S\,{\sc ii}]/H$\alpha$ and [N\,{\sc ii}]/H$\alpha$ ratios; hence the contribution from shocks cannot be totally ignored. 

A similar effect is found in the [O\,{\sc iii}]\,5007/H$\beta$ and [O\,{\sc iii}]\,5007/H$\alpha$ ratios. The values are higher in the regions closer to the central ionizing stars showing a high degree of ionization, while the ratios are lower at the outer regions indicating a low degree of ionization. 
We note that, the integrated [O\,{\sc iii}]\,5007/H$\alpha$ ratio of N44\,D1 (2.52) is much higher than N44\,D2 (1.67) and N44\,C (0.55), which indicates the hardness of radiation field in N44\,D1. The [O\,{\sc iii}]\,5007/H$\beta$ ratio is also higher in N44\,D1, representing the effect of high effective temperature of the ionizing star of spectral type O5\,V. The N44\,D1, N44\,D2, and N44\,C show an [O\,{\sc iii}]\,5007/H$\beta$ ratio of 7.89, 5.18 and 2.07 respectively. In N44\,C, the high values of [O\,{\sc iii}]\,5007/H$\alpha$ ($\sim$\,1.0 -- 0.3) and [O\,{\sc iii}]\,5007/H$\beta$ ($\sim$\,3.0 -- 1.0) are near the O5\,III star and the values decrease toward the edge of the bubble.

\section{H$\alpha$ and H$\beta$ emission}
\subsection{Extinction correction}

The line luminosities are corrected for extinction using the intensity ratios (H$\alpha$/H$\beta$)$_{\textrm{obs}}$. Since, H$\alpha$/H$\beta$ ratio is relatively sensitive to temperature, it can be used as a reliable reddening indicator. This ratio is compared with the theoretically expected value of the Balmer decrement (H$\alpha$/H$\beta)_{\textrm{exp}}$ for Case B recombination \citep{Osterbrock2006}.
Any deviation from expected value of H$\alpha$/H$\beta$ ratio for a particular electron temperature can be associated with extinction. We estimate the nebular emission line color excess $E(B-V)$ from the H$\alpha$/H$\beta$ ratio using the equation from \citet{Dom_nguez_2013},
\begin{equation}
E(B-V) = \frac{2.5}{k(\lambda_{\textrm{H}\beta})-k(\lambda_{\textrm{H}\alpha})} log_{10}\Bigg[\frac{({\textrm{H}\alpha}/{\textrm{H}\beta})_{\textrm{obs}}}{({\textrm{H}\alpha}/{\textrm{H}\beta})_{\textrm{exp}}}\Bigg]
\end{equation}

The expected value of (H$\alpha$/H$\beta)_{\textrm{exp}}$ flux ratio is $\sim$\,2.86 \citep{Osterbrock2006}. This value is obtained by assuming Case B recombination at an electron temperature 10\,000\,K and density 100\,cm$^{-3}$. Then, following the extinction curve estimated by \citet{calzetti2000}, 

\begin{equation}
k(\lambda)=2.659(-1.857+1.040/\lambda)+ R_{\textrm{V}}  
\end{equation}

for $\lambda$ = 0.63 $\mu m$ to 2.2 $\mu m$, and
\begin{equation}
k(\lambda)=2.659(-2.156+1.509/\lambda - 0.198/\lambda^2 +0.011/\lambda^3)+ R_{\textrm{V}}
\end{equation}
for $\lambda$ = 0.12 $\mu m$ to 0.63 $\mu m$.\\

Here, $k(\lambda_{\textrm{H}\alpha})$ and $k(\lambda_{\textrm{H}\beta})$ are the extinction curves at H$\alpha$ and H$\beta$ wavelengths respectively.

Assuming the ratio of total to selective extinction $R_\textrm{V} (=A_{\textrm{V}}/E(B-V))=3.1$, which is valid at optical wavelength for the LMC \citep{Gordon03}, we get $k(\lambda_{\textrm{H}\alpha})$ = 2.38 , and $k(\lambda_{\textrm{H}\beta})$ = 3.65 . 

Using the color excess $E(B-V)$, the extinction in magnitude for H$\alpha$ line is obtained by
\begin{equation}
    A_{\textrm{H}\alpha} = k(\lambda_{\textrm{H}\alpha}) E(B-V) 
\end{equation}
and for H$\beta$ line,
\begin{equation}
A_{\textrm{H}\beta} = k(\lambda_{\textrm{H}\beta}) E(B-V) 
\end{equation}
Then extinction corrected H$\alpha$ luminosity, $L$(H$\alpha)$ is,
\begin{equation}
    L(\textrm{H}\alpha) =  L(\textrm{H}\alpha)_{obs} 10^{0.4 A_{\textrm{H}\alpha}}
\end{equation}
and the extinction corrected H$\beta$ luminosity, $L$(H$\beta)$ is,
\begin{equation}
    L(\textrm{H}\beta) =  L(\textrm{H}\beta)_{obs} 10^{0.4 A_{\textrm{H}\beta}}
\end{equation}


Here $L$(H$\alpha)_{obs}$ and $L$(H$\beta)_{obs}$ are the observed luminosities of H$\alpha$ and H$\beta$ emission respectively. 

The values of the color excess $E(B-V)$ are 0.08 for N44\,D1 and N44\,D2, and  0.20 for N44\,C. The extinction toward N44\,D1 is $A_{\textrm{V}} = 0.25$\,mag. Our value agrees with the calculations by \citet{Garnett_2000} and \citet{Lopez_2014} for N44. The N44\,C has a significantly higher extinction, $A_{\textrm{V}} = 0.62$\,mag.

We applied the same method of extinction correction to other line emission, and the extinction corrected luminosities (erg s$^{-1}$) are given in Table\,\ref{lines}. 


\subsection{Lyman Continuum Photon Flux}
The O-type stars in H\,{\sc ii} regions are the prominent sources of Lyman continuum photons. These stars deposit a bulk of their high energy photons into the surrounding H\,{\sc ii} region within the Stromgren radii. If the gas is optically thick in the Lyman continuum, we expect all the ionizing photons emitted by the star to be absorbed. However, a significant fraction of these photons can escape on a larger scale outside of the H\,{\sc ii} region into the ISM. This fraction of photon leakage from H\,{\sc ii} regions needs to be measured to understand the overall energy budget, and to probe whether the dominant hot stars in the region are responsible for the photoionization in the surrounding medium. We calculate the number of L[OIII]/H$\beta$yman continuum photons ($Q$) absorbed in the region surrounding a hot star corresponding to H$\alpha$ luminosities by assuming the Case B recombination for electron temprature $T_{\textrm{e}}$ = 10\,000\,K, and density $n_{\textrm{e}}$ = 100\,cm$^{-3}$. The number of Lyman continuum photon related to H$\alpha$ luminosity is obtained by $Q(\textrm{H}\alpha)$ = 7.31$\times$10$^{11}L$(H$\alpha$)\,ph\,s$^{-1}$ \citep{kennicutt1998, Osterbrock2006}.

The number of ionizing photons derived from the H$\alpha$ luminosity ($Q$) for N44\,D1, N44\,D2, and N44\,C are tabulated in Table\,\ref{properties}.
To calculate the photon escape fraction, we also need to know the number of total Lyman continuum photons emitted by the ionizing stars. We adopt the model calculations ($Q_0$) for hot stars of appropriate spectral types from \citet{martins05}. The $Q_0$ values of the only O5\,V star in N44\,D1, a combination of three ionizing stars of spectral types O5\,III, O8\,V and O9.5\,V in N44\,C, and O5.5\,V star in N44\,D2 are given in Table\,\ref{properties}. 

Using the $Q$ and $Q_0$ values we calculate the photon escape fraction,

\begin{equation}
f_{\textrm{esc}} = \frac{Q_0-Q}{Q_0} 
\end{equation}

\citet{McLeod19} reported $f_{\textrm{esc}} \backsim$ 0.37 and 0.68 for N44\,D1 and N44\,C respectively, using this method. We verify their determination and find that $f_{\textrm{esc}}$ for N44\,D1, N44\,D2, and N44\,C are 0.36, 0.71, and 0.70 respectively. These values imply that about 36$\%$ of the ionizing photons escape from N44\,D1, 71$\%$ of ionizing photons escape from N44\,D2 and 70$\%$ from N44\,C. N44\,C shows a relatively larger amount of photon leakage than N44\,D1, and is more optically thin to the ionizing photon. This observation is consistent with the study of H\,{\sc ii} regions in the LMC by \citet{pellegrini12} that boundaries of optically thick regions are generally characterized by stratification in the ionization structures. The ionization structure of N44\,D1 shows a well-defined nebular boundary where [O\,{\sc i}] and [S\,{\sc ii}] emission peak at the outer boundary of the H$\alpha$ and [O\,{\sc iii}] emission. The [S\,{\sc ii}]/H$\alpha$ ratio map clearly shows the transition region in the ionization structure. N44\,C does not show such an ionization stratification, while a slightly extended shell structure is found in the [S\,{\sc ii}]/H$\alpha$ map. 

Using the number of ionizing photons $Q$, we can also estimate the average electron density of emitting gas in an H\,{\sc ii} regions, $\langle n_{\textrm{e}}\rangle$. Assuming the spherical nebula where H is fully ionized, the recombination balance equation is,

\begin{equation}
Q=\frac{4\pi}{3} \alpha_{\textrm{B}} n_{\textrm{e}}^2 R^3_{\textsc{Hii}}
\end{equation}

Here $\alpha_{\textrm{B}}$ is the Case B recombination coefficient $\sim$ 2.59$\times$10$^{-13}$ cm$^{3}$ s$^{-1}$ for gas at $T$\,=\,10\,000\,K. $R_{\textrm{H{\sc ii}}}$ is the mean radius of t[OIII]/H$\beta$he H\,{\sc ii} region. Then the average electron density from H$\alpha$ emission is obtained by,

\begin{equation}
\langle n_{\textrm{e}}\rangle\, = 177 \sqrt{\frac{Q_{48}}{R^3_{\textsc{Hii}}}}
\end{equation}

Here $Q_{48}$(=Q/10$^{48}$) is the number of Lyman continuum photons derived from the H$\alpha$ luminosity and $R_{\textrm{H{\sc ii}}}$ is the radius in parsec.
The $\langle n_{\textrm{e}}\rangle$ derived from H$\alpha$ emission of N44\,D1, N44\,D2, and N44\,C are 31, 26, and 38 cm$^{-3}$ respectively.
\subsection{Electron density}
Electron density ($n_{\textrm{e}}$) and electron temperature ($T_{\textrm{e}}$) are two important physical parameters for characterizing an H\,{\sc ii} region. The $n_{\textrm{e}}$ can be determined from the observed line intensities of two different energy levels with nearly equal excitation energy of the same ion. Their line ratios are generally not sensitive to $T_{\textrm{e}}$. The forbidden lines ratio, [S\,{\sc ii}]\,6717/6732 is usually used to determine $n_{\textrm{e}}$, where [S\,{\sc ii}]\,6717\AA\, and 6732\AA\, emission are relatively strong in the ionized nebula. Their corresponding critical density is $\sim$\,10$^{3}$\,cm$^{-3}$, hence probing the low-density regimes. \citet{McLeod19} have reported $n_{\textrm{e}}$\,$\sim\,152\pm42$\,cm$^{-3}$ for N44\,C and $\sim\,143\pm42$\,cm$^{-3}$ for N44\,D1, applying the analytical solution given in \citet{McCall84} and assuming a $T_{\textrm{e}}$ of 10\,000\,K. \citet{Toribio17} have derived  $n_{\textrm{e}}\,\sim\,200\pm150$\,cm$^{-3}$ for a 3.0$\times$9.4 arcsec$^2$ region closer to the ionizing star in N44\,D1 using [S\,{\sc ii}]\,6717/6732 ratio obtained from VLT-UVES spectrum. \citet{Lopez_2014} have reported a relatively low value of $n_{\textrm{e}}$\,$\sim\,60$\,cm$^{-3}$ for the entire N44 using the flux density of the free-free emission at 3.5 cm. \citet{Garnett_2000} derived $n_{\textrm{e}}$\,$ <\,160\,cm^{-3}$ for N44D1 using [S\,{\sc ii}]\,6717/6732 ratio obtained with 0.9 m telescope at Cerro Tololo Inter-American Observatory. \citet{McLeod19} emphasize that, densities derived from radio emission by \citet{Lopez_2014} are expected to be smaller than those derived from the ratio of collisionally excited lines \citep{peimbert2017}. These studies show a discrepancy in derived values of $n_{\textrm{e}}$ for N44\,D1 and N44\,C, hence we calculate the electron densities using [S\,{\sc ii}]\,6717/6732 ratios of N44\,D1 and N44\,C derived in our analysis of MUSE spectra. Electron temperature can be obtained by the forbidden line ratios [SIII]6312/9069 and [NII]5755/6384, however the MUSE observations of N44\,D1 and N44\,C show very weak [SIII]\,6312\AA\, and [NII]\,5755\AA\, emission, which cannot be extracted from the data cube within a 5$\sigma$ detection threshold in most of the pixels inside the defined polygons. We calculate the electron density as in \citet{McLeod2015} by applying the analytical solution in \citet{McCall84}.

\begin{equation}
    n_e = \frac{1.49-R_{\textsc{sii}}}{12.8\times R_{\textsc{sii}}- 5.6713}\times10^4 \;\;\mbox{cm}^{-3}
\end{equation}

Here, R$_{\textsc{sii}}$ is the [S\,{\sc ii}]\,6717/6732 ratio, and electron temperature is assumed to be 10\,000\,K as in \citet{McLeod19}. The derived $n_{\textrm{e}}$ for N44D1, N44D2 and N44C using this method are 141$\pm$43, 121$\pm$37 and 92$\pm$35 \,cm$^{-3}$ respectively. We also estimate $n_e$ using the publicly available Python based package, PYNEB \citep{Luridiana2013} for a comparison. This algorithm includes FIVEL \citep{robertis1987} and NEBULAR \citep{shaw1995} packages for analyzing nebular emission lines. The package calculates the physical conditions ($T_{\textrm{e}}$ and $n_{\textrm{e}}$) for a given set of emission line intensities, and returns the diagnostic plots. We use the density sensitive [S\,{\sc ii}]\,6732/6717 line ratio to determine $n_{\textrm{e}}$ using the diags.getTemDen task in PYNEB for a given $T_{\textrm{e}}$ of 10\,000\,K. The estimated electron densities from PYNEB for N44D1, N44D2 and N44C are 132$\pm$50, 115$\pm$45 and 66$\pm$40 \,cm$^{-3}$ respectively (Table\,\ref{properties}). These density values are comparable with the density derived from the equation 11 within the estimated uncertainties.

\section{Structure of ionized gas}
\begin{figure*}
    \centering
    \includegraphics[scale=0.35]{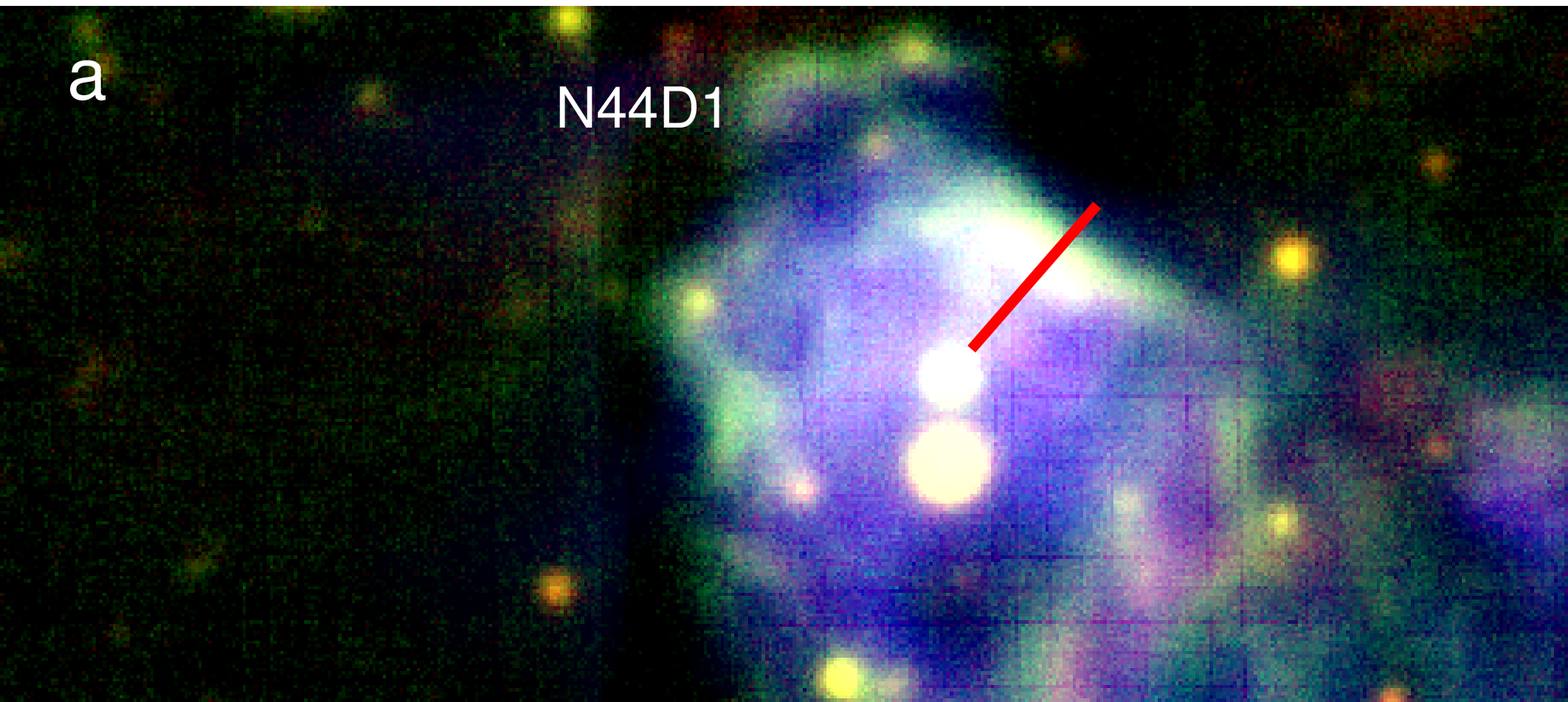}\\
    \includegraphics[scale=0.35]{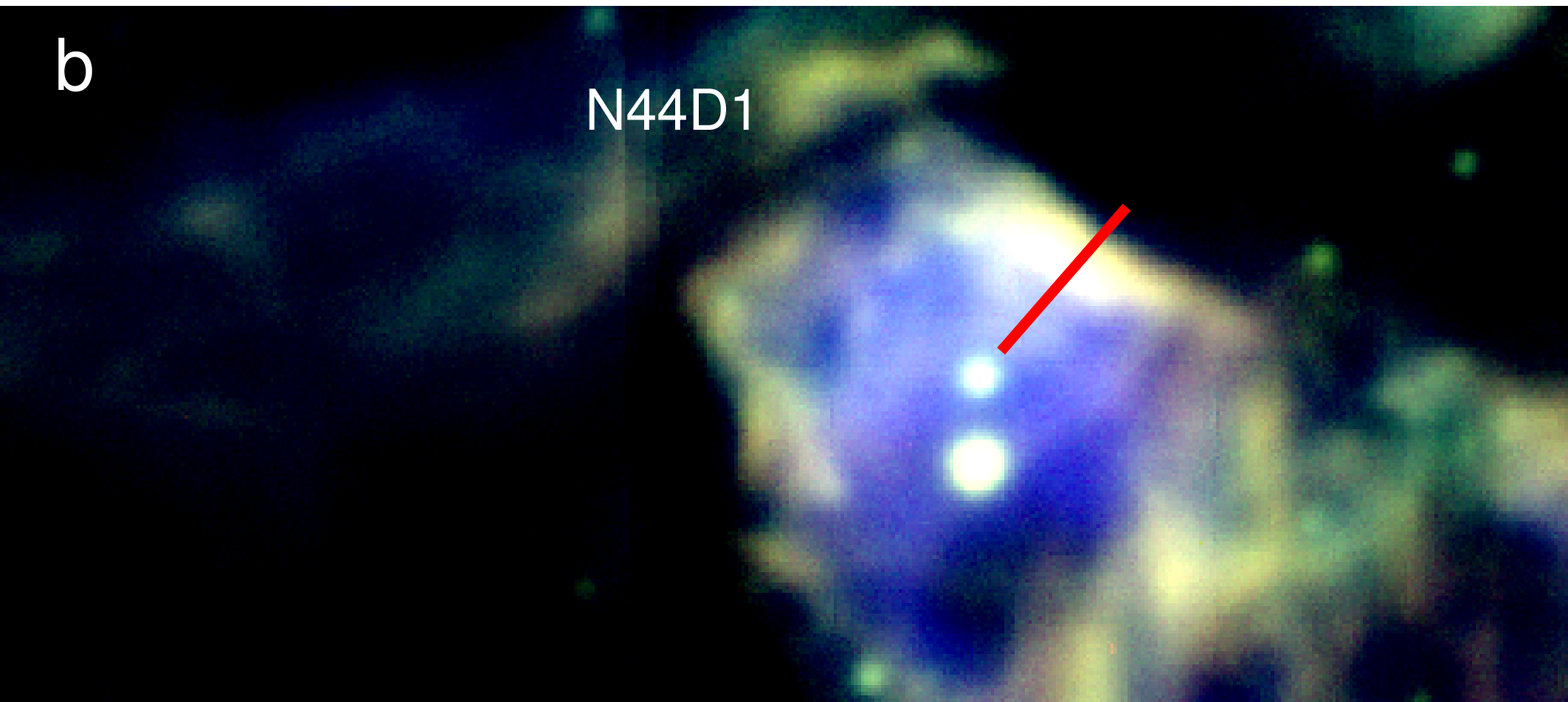}\\
    \includegraphics[scale=0.35]{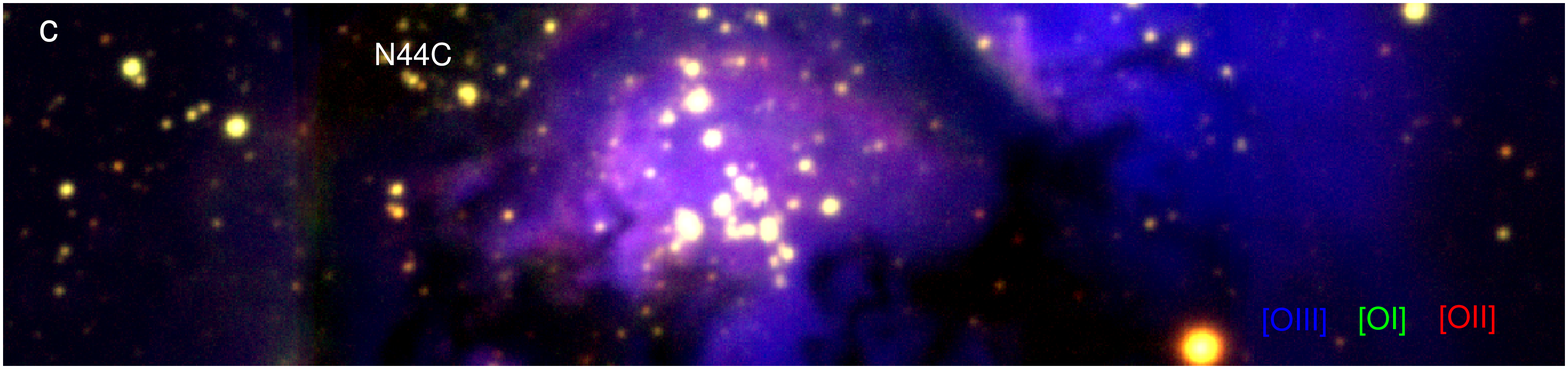}\\
    \includegraphics[scale=0.35]{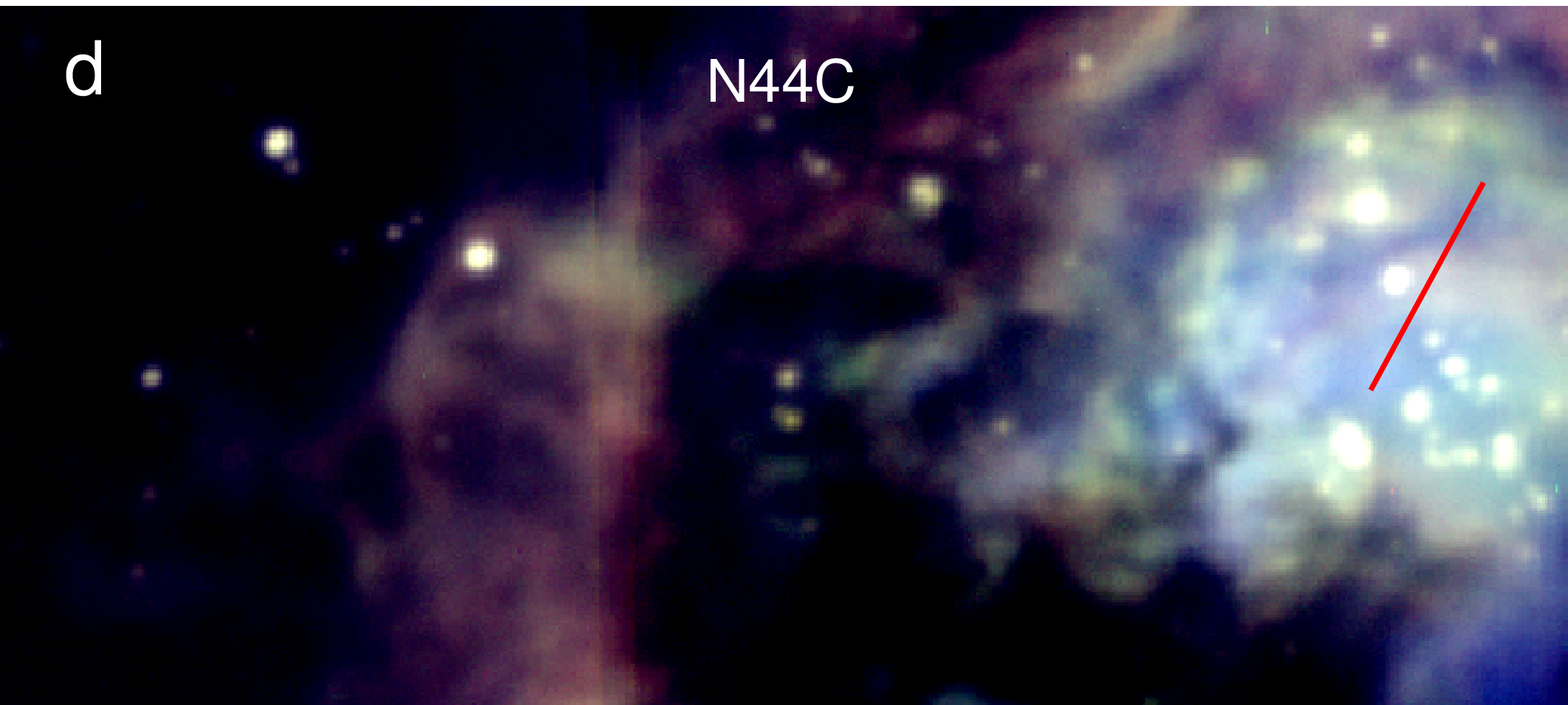}
    \caption{(a) Spatial distribution of N44\,D1 and N44\,D2 are shown in [O\,{\sc iii}]\,5007\AA\, (blue), [O\,{\sc ii}]\,7318\AA\, (red) and [O\,{\sc i}]\,6300\AA\, emission (green). (b) N44\,D1 and N44\,D2 are shown in H$\beta$ (blue), [S\,{\sc ii}]\,6717\AA\, (red) and [N\,{\sc ii}]\,6584\AA\, emission (green). (c) N44\,C is shown in [O\,{\sc iii}]\,5007\AA\, (blue), [O\,{\sc ii}]\,7318\AA\, (red) and [O\,{\sc i}]\,6300\AA\, emission (green). (d) N44\,C is shown in H$\beta$ (blue), [S\,{\sc ii}]\,6717\AA\, (red), and [N\,{\sc ii}]\,6584\AA\, (green). The ionization zone is traced by [O\,{\sc iii}] and H$\beta$, the partially ionized zone is traced by [O\,{\sc ii}] and [N\,{\sc ii}] and the ionization front is traced by [O\,{\sc i}]\,6300\AA\, and [S\,{\sc ii}]\,6717\AA\,. The observed spatial profiles in Fig. \ref {fig:spatialcut1} and \ref{lineprof_mod_N44D} are taken along the cross-cuts indicated as red lines in these maps.}
    \label{fig:spatialmap}
\end{figure*}

To investigate the structure of ionization zones in N44\,D and N44\,C we compare the spatial distribution of H$\beta$, [O\,{\sc iii}], [O\,{\sc ii}], [O\,{\sc i}], [N\,{\sc ii}] and [S\,{\sc ii}] emission line maps (Fig.\,\ref{fig:spatialmap}). The spatial distribution of [O\,{\sc iii}], [O\,{\sc ii}], and [O\,{\sc i}] of N44\,D1 in Fig.\,\ref{fig:spatialmap}(a) shows a clear stratification from ionization zones where [O\,{\sc iii}]\,5007\AA\, emission peaks around the O5\,V star, to the [O\,{\sc i}]\,6300\AA\, emission at the ionization front. \citet{odell1992} have reported that [O\,{\sc i}]\,6300\AA\, emission arises at the ionization front behind the PDR, between the ionization zone and molecular cloud. The majority of the [O\,{\sc i}]\,6300\AA\, emission in H\,{\sc ii} regions is due to the collisional excitation by thermal electrons and atomic hydrogen via the charge exchange, hence the intensity of [O\,{\sc i}]\,6300\AA\, emission can be a measure of neutral hydrogen content. [O\,{\sc i}]\,6300\AA\, in PDRs can also be collisionally excited by electrons ejected by dust grains and Polycyclic Aromatic Hydrocarbon (PAHs) that absorb far-UV radiation emitted by the massive stars. In Fig.\,\ref{fig:spatialmap}(b), we compare the spatial distribution of H$\beta$ (blue) and [N\,{\sc ii}]\,6584\AA (red) with [S\,{\sc ii}]\,6717\AA (green). [S\,{\sc ii}] emission is expected to peak at ionization front. [N\,{\sc ii}] emission is found to be co-spatial with [O\,{\sc ii}] emission that peaks at partial-ionization zone, and concentrated in the outer boundary of the H$\beta$ and [O\,{\sc iii}] emission. The structure of N44\,D1 is nearly spherical with only one important source of ionizing photons.

\begin{figure}
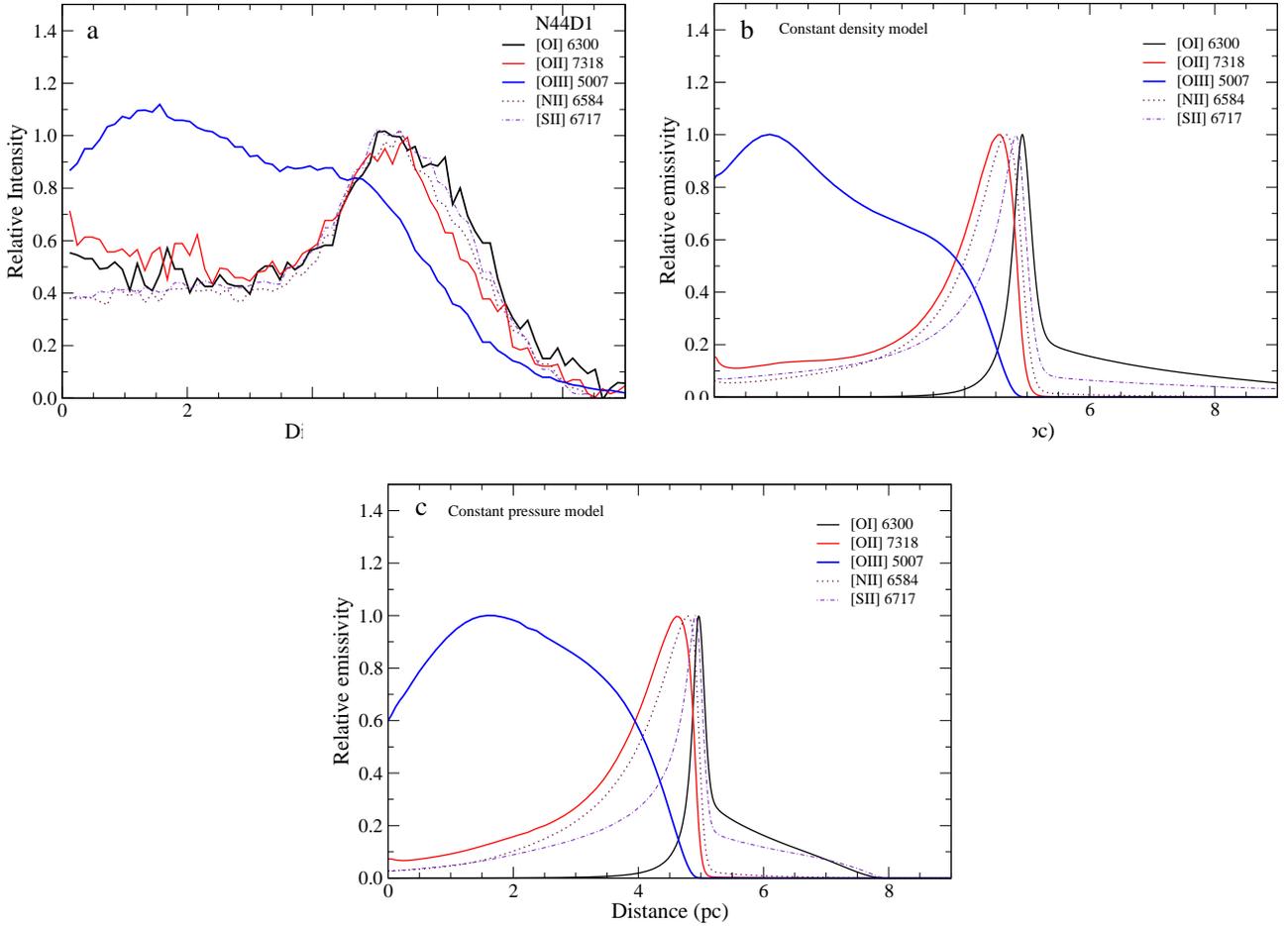

    \centering
    \includegraphics[scale=0.35]{OIII_OII_OI_profile4_grace_pc_latest.eps}
    \hspace{0.2 cm}
    \includegraphics[scale=0.35]{N44D_constantdensity_spatialprof_mod_grace.eps}\\
    \vspace{0.5 cm}
    \includegraphics[scale=0.35]{N44D_constantpressure_spatialprof_grace_mod.eps}
    \caption{ The spatial profiles of various emission lines in MUSE observations of N44\,D1 are shown for comparison with the constant density model and constant pressure model. a) The observed spatial profiles of [O\,{\sc i}]\,6300\AA, [O\,{\sc ii}]\,7318\AA, [O\,{\sc iii}]\,5007\AA, [N\,{\sc ii}]\,6584\AA, and [S\,{\sc ii}]\,6717\AA \,emission of N44\,D1 obtained by taking a cross-cut (red line) shown in Fig.\,\ref{fig:spatialmap}. b) The emission line profiles of [O\,{\sc i}]\,6300\AA, [O\,{\sc ii}]\,7318\AA, [O\,{\sc iii}]\,5007\AA, [N\,{\sc ii}]\,6584\AA, and [S\,{\sc ii}]\,6717\AA \, emission from constant density model of N44\,D1 are shown for comparison. c) The spatial profiles of [O\,{\sc i}]\,6300\AA, [O\,{\sc ii}]\,7318\AA, [O\,{\sc iii}]\,5007\AA, [N\,{\sc ii}]\,6584\AA, [S\,{\sc ii}]\,6717\AA \,emission from constant pressure model of N44\,D1 are shown for comparison. The best-fit model line ratios for these models are given in Table  \ref{linesratio_N44D1}. All the line emissivities are normalized to 1.0.}
    \label{fig:spatialcut1}
\end{figure}

In Fig.\,\ref{fig:spatialcut1}  we show the spatial profiles of various nebular emission lines by taking a cross-cut along the north-western edge of N44\,D1. [O\,{\sc ii}] and [N\,{\sc ii}] emission are in a thin layer at the partial ionization zone outside [O\,{\sc iii}], but slightly interior to the [O\,{\sc i}] emission. The [O\,{\sc i}] emission is concentrated in a thin zone $\sim$\,4 -- 7\,pc located in the outer boundary of the H\,{\sc ii} region. H$\alpha$, H$\beta$, [O\,{\sc iii}] and He\,{\sc i} are found to be co-spatial in the ionization zone. In the outer layer of the ionization front in this H\,{\sc ii} region, we expect a well-defined PDR with a layer of C\,{\sc ii}, then H$_2$ emission, and a molecular cloud. High-spatial resolution spectroscopic observations in infrared and submillimeter wavelengths are required for further studies of PDR properties.  A similar ionization structure is reported in the Orion Nebula H\,{\sc ii} region. The [O\,{\sc i}]\,6300\AA\, and [S\,{\sc ii}]\,6717\AA\, emission appear to peak along a bright bar at the ionization front in the outer boundary, that forms a thin transition layer of thickness $\sim$\,10$^{15}$--10$^{16}$ cm between the ionization zone and PDR.The emission from higher ionization species [O\,{\sc iii}] arises away from the ionization front close to the ionizing star $\theta_1$Ori C \citep{Odell2017, Odell01, Hester1991, odell1992}. 

In N44\,C, the [O\,{\sc iii}]\,5007\AA\, and H$\beta$ emission appear to peak in the interior of the bubble near the ionizing star O5\,III (Figs.\,\ref{fig:spatialmap} c, d). We show the spatial profiles of [O\,{\sc iii}], [O\,{\sc ii}], [S\,{\sc ii}], [N\,{\sc ii}], and H$\alpha$ emission in N44\,C taking a cross-cut from the position centroid of three ionizing stars to the north-east of N44\,C (Fig.\,\ref{fig:spatialcut1}). The spatial distribution of [O\,{\sc iii}]\,5007\AA\, and H$\beta$, are similar and co-spatial in the fully ionized zone. [O\,{\sc i}]\,6300\AA, [O\,{\sc ii}]\,7318\AA, [N\,{\sc ii}]\,6584\AA\, and [S\,{\sc ii}]\,6717\AA\, emission do not form a well-defined outer boundary, but appear as patches of emission bars within [O\,{\sc iii}]\,5007\AA\, and H$\beta$ emission region. 

\section{Comparison with photoionization model}

The MUSE observations of N44 in the LMC show that the two bright H\,{\sc ii} regions N44\,D1 and N44\,C have different ionization geometries and physical characteristics. N44\,D1 has H$\alpha$ and H$\beta$ surface brightness values 0.03\,dex and 0.1\,dex higher than N44\,C. 
[O\,{\sc iii}]/H$\beta$ and [O {\sc iii}]/H$\alpha$ ratios in N44\,D1 are considerably larger than in N44\,C, indicating a higher degree of ionization. N44\,D1 shows a higher [O\,{\sc i}]\,6300/H$\beta$ ratio with a well-defined [O\,{\sc i}]\,6300\AA\,emission at the H\,{\sc ii} region outer boundary, indicating an ionization front. 

 In the ideal case, the [O\,{\sc i}]\,6300\AA\, emission dominates at the outer boundary of the H\,{\sc ii} region where the neutral hydrogen density dominates; hence we expect most of the Lyman continuum photons are absorbed by the nebula is ionization-bounded. This is an optically thick case, where we find a shell of [O\,{\sc i}]\,6300\AA\, emission at the ionization front, indicating the border of the H\,{\sc ii} region. In the density-bounded H\,{\sc ii} regions, there is no well-defined ionization front surrounding highly ionized gas, and most of the Lyman continuum photons leak from the cloud contributing to the diffuse ionized gas outside of the cloud. In such cases, there is a weak [O\,{\sc i}]\,6300\AA\, emission condensation within the H\,{\sc ii} region or no shell structure at the boundary. In a Blister H\,{\sc ii} region, there is a partial ionization front at the boundary, which does not cover the nebula completely; hence the photons escape in certain directions. Our study of the [O\,{\sc i}]\,6300\AA\, emission in N44\,D1 and N44\,C reveals these different observational properties of H\,{\sc ii} regions. Moreover, the measurement of the photon leakage using the H$\alpha$ emission indicates that N44\,D1 and N44\,C have photon escape fractions 36$\%$ and 70$\%$, respectively. These calculations are in agreement with the study of the photon leakage by \citet{McLeod19}. The remaining ionizing photons are trapped within the H\,{\sc ii} region itself and affect the overall ionization balance. \citet{pellegrini11,pellegrini12} reported that the ionization-bounded H\,{\sc ii} regions are constrained to an escape fraction $<$\,0.6 and those with density-bounded are $>$\,0.6. 
 
To further interpret these observations, we compute various photo-ionization models for comparing the emission line ratios and the geometry of ionization structure.
We model the ionization structure of N44\,D1 and N44\,C using the photoionization code CLOUDY \citep{ferland2017}. We derive the emissivities of prominent ionic species across the H\,{\sc ii} region, from the illuminated face of the H$^+$ region through the partially ionized zone to the neutral ionization front, where ionizing radiation has been attenuated and becoming neutral to the molecular zone. We develop various photoionization models for N44\,D1 and N44\,C, and test which model can better match the observed ionization geometry of the cloud and emission line ratios. To compare the ionization geometry, we use the spatial profiles of the line emissions along the H\,{\sc ii} regions (Figs.\,\ref{fig:spatialcut1} and \ref{lineprof_mod_N44D}). The emission line ratios we use for tests are given in Tables \ref{linesratio_N44D1} and \ref{linesratio_N44C}. The [S\,{\sc ii}]\,6717/6732 line ratio is sensitive to electron density, and [O\,{\sc ii}]\,(7318+7329)/[O\,{\sc iii}]\,5007 to the ionization parameter. [O\,{\sc iii}]\,5007/H$\alpha$, [O\,{\sc iii}]\,5007/H$\beta$, [O\,{\sc ii}]/H$\beta$, [N\,{\sc ii}]/H$\beta$, and [O\,{\sc i}]/H$\beta$ give the behavior of different ionization zones, and are also sensitive to the metallicity. These line ratios are very sensitive to the adopted input parameters. A built-in optimization program based on the PHYMIR algorithm \citep{vanhoof97, vanhoof13, Ferland13} is used to obtain the best fit model, which applies a $\chi^2$ minimization to determine the goodness-of-fit by varying the input parameters. We vary the total hydrogen density ($hden$), $\Phi$(H), $brems$ and filling factor to find the best agreement between the model with the observed line ratios and geometry. Sometimes, for a given set of constraints some observables are optimized very well compared to the other sets and the best-fit model is obtained by the overall $\chi^2$. Therefore, we manually fine tuned certain parameters until the best matching line ratios with the observations are obtained. Finally, the model line ratios are compared with the observed line ratios and the best-fit model was determined by calculating the $\chi ^2$ as \citep{mondal2017, pavana_2019},

 
 
\begin{equation}
\chi^2 = \sum_{i=1} ^{n}(M_i - O_i)^2/\sigma_i ^2
\end{equation}
Here, the number of observed lines is n, M$_i$ is the model line ratio, O$_i$ is the observed line ratio and the $\sigma_i$ is the error in the observed flux ratio. The best optimized model line ratios, observed line ratios and the $\chi^2$ values are given in Tables \ref{linesratio_N44D1} and \ref{linesratio_N44C}. 




The basic input parameters to CLOUDY require the geometry of the cloud, intensity of incident ionizing photon flux, elemental abundances and gas density.  
\subsection{Geometry}
Observations show a clear ionization stratification in N44\,D1. The ionized gas is traced by [O\,{\sc iii}]\,5007\AA\, and H$\beta$ emission and the ionization front is traced by [O\,{\sc i}]\,6300\AA\, and [S\,{\sc ii}]\,6717\AA\, emission. The line ratio maps show a nearly radial symmetry around the ionizing star. This structure is quite simple to model as observation shows there is only a prominent source of the ionizing photon at the cloud's interior. For N44\,D1, we calculate an optically thick spherical model with a covering factor 0.64. We choose an outer radius, $R$ = 7.4\,pc based on the observed geometry (Fig.\,\ref{fig:spatialcut1}). \citet{McLeod19} report a radius containing 90 per\,cent of the H$\alpha$ emission, $R_{90}$ = 7.4\,pc for N44\,D1 which agrees with the adopted radius in our model. The model constitutes the exciting star at the center of a spherical cloud, surrounded by the layers of H\,{\sc ii} region and PDR. 

MUSE observation of N44\,C does not show a clear ionization stratification as N44\,D1. However, observations indicate that the majority of photons escape from N44\,C and the region harbors three ionizing stars (See Fig.\,\ref{N44_OB}). Therefore, we adopt an optically thin open geometry for N44\,C with a covering factor 0.3 and a radius of $R$ = 6.93\,pc.
\subsection{Stellar Continuum}
We use OSTAR TLUSTY models in CLOUDY for defining the stellar continuum. 
We choose a model with the effective temperature ($T_{\textrm{eff}}$) of 41\,540\,K, gravity (log$g$) of 3.92, and a metallicity of 0.5Z$_\odot$, that is appropriate for a spectral class O5\,V star \citep{McLeod19} in N44\,D1. The radiation field from this O5\,V star corresponds to an incident flux of ionizing photons, log\,$\Phi(H) = 9.98\,$ph$\,s^{-1}$\,cm$^{-2}$ at the ionization front of N44\,D1 traced by the peak of [O\,{\sc i}]\,6300\AA\, emission. Here, $\Phi(H) = Q(\textrm{H})/4\pi r^2$, where the number of ionizing photons per second $Q(\textrm{H})$ is 1.82$\times10^{49}$\,ph\,s$^{-1}$ and $r$ is the distance from ionizing star to the cloud illumination face. We note that the flux of ionizing photons $\Phi$(H), plays a significant role in shape of spatial profiles and lines strengths; hence we test the models with varying $\Phi$(H) until the best-fit model is obtained. For N44\,D1 we vary the $\Phi$(H) in a range 9.98$-$10.60\,ph\,s$^{-1}$\,cm$^{-2}$. 

Observations show that, N44\,C encloses three hot stars of spectral types O5\,III, O8\,V, and O9.5\,V, hence we chose three hot star model atmospheres from TLUSTY. We adopt models with $T_{\textrm{eff}}$ = 39\,500\,K and log$g$ = 3.69 for spectral type O5\,III, $T_{\textrm{eff}}$ = 33\,400\,K and log$g$ = 3.92 for O8\,V star, and $T_{\textrm{eff}}$ = 30\,500\,K and log$g$ = 3.92 for O9.5\,V star with a metallicity of 0.5\,Z$_{\odot}$.

 We expect the X-rays to significantly affect the ionization balance of the gas, since the observed total X-ray luminosity is in comparable range as that of the observed H$\alpha$ luminosity in both N44\,D1 and N44\,C. \citet{chu1993} have presented the global X-ray emission in N44 using observations with the ROSAT satellite. They reported a diffuse X-ray luminosity of (0.29$-$3.5)$\times 10^{37}$ erg\,s$^{-1}$ at a characteristic temperature (1.6$-$2.5)$\times 10^{6}$\,K. We vary the Bremsstrahlung temperature ($brems$) in a range (1.6$-$2.5)$\times 10^{6}$ K.
\subsection{Abundances}
We follow the abundances provided by \citet{Toribio17} for C and O, and \citet{Garnett_2000} for He, N, Ne, S, and Ar. For the remaining species, we use the standard values included in CLOUDY for H\,{\sc ii} regions \citep{Baldwin91} and adopt an overall gaseous metallicity of 0.5\,Z$_{\odot}$. Since the dust contributes significantly to heating and the overall equilibrium of the photoionized gas in the cloud, we also include the dust grains with a metallicity scaled to half solar. 

\subsection{Density}
For gas density at the illuminated face of the cloud, we use the range of electron density values obtained from the H$\alpha$ luminosity and [S\,{\sc ii}]\,6717/6732 line ratio (30 $\le n_{\textrm{e}} \le$ 180)\,cm$^{-3}$ for N44\,D1. For N44\,C, we varied the density between 40$-$100\,cm$^{-3}$, and finally obtained a value consistent with the observed electron density obtained from [S\,{\sc ii}]\,6717/6732 ratio (66\,cm$^{-3}$). We compute models with constant density as well as constant pressure in a time-steady hydrostatic cloud. When we use a filling factor of 1, we find most of the predicted line ratios differ from the observed values, and the ratios considerably change when we use a filling factor $<$\,1. We estimate an approximate range of filling factors using the relation $N^{2}_{\textrm{e}}$(rms) = $\epsilon$ $N^{2}_{\textrm{e}}$(local) \citep{Relano_2002}. Here, $N_{\textrm{e}}$(rms) is taken as the average electron density derived from H$\alpha$ flux and $N_{\textrm{e}}$(local) is the electron density obtained from the [S\,{\sc ii}]\,6717/6732 ratio.

We note that the total gas and radiation pressure vary with the position and width of the ionization front. This can also be a result of varying density at the ionization front as pressure changes. If density increases, the ionization front pushes the interior to the cloud, increasing the line emissivities. We therefore, tested two models: one with constant pressure and the other with constant density distribution. 

\subsection{Constant pressure and constant density distribution}

\begin{figure*}
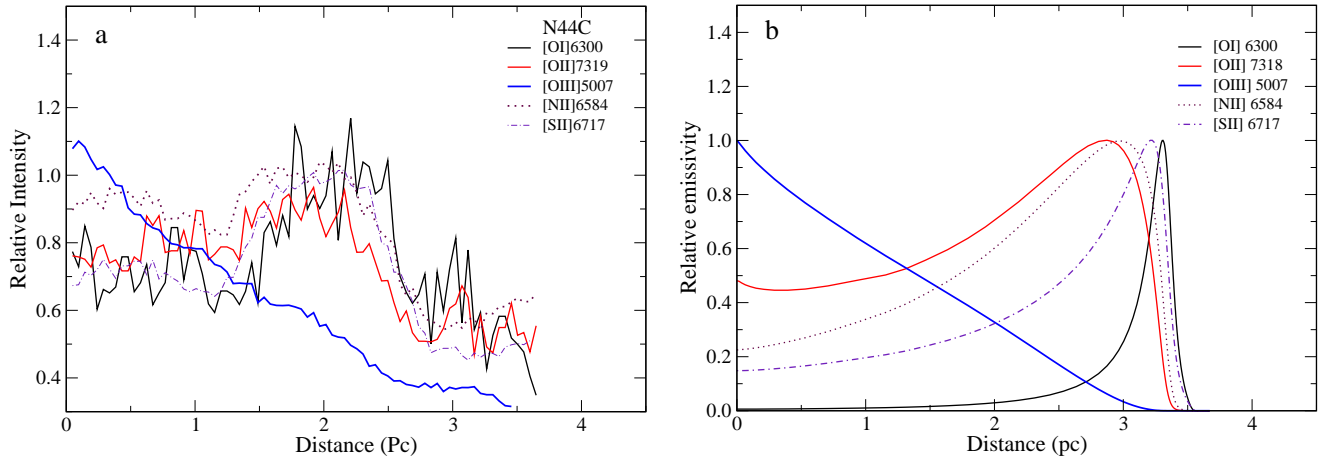

    \includegraphics[width=8.5cm]{N44C_2-spatial_prof_grace_latest.eps}
    \hspace{0.2 cm}
    \includegraphics[width=8.5cm]{N44C_constantdensity_spatial_prof_grace.eps}
    \caption{ a) The observed spatial profiles of [O\,{\sc i}]\,6300\AA, [O\,{\sc ii}]\,7318\AA, [O\,{\sc iii}]\,5007\AA, [N\,{\sc ii}]\,6584\AA, and [S\,{\sc ii}]\,6717\AA \,emission of N44\,C obtained by taking a cross-cut (red line) shown in Fig.\,\ref{fig:spatialmap}. b) The spatial profiles of [O\,{\sc i}]\,6300\AA, [O\,{\sc ii}]\,7318\AA, [O\,{\sc iii}]\,5007\AA, [N\,{\sc ii}]\,6584\AA, and [S\,{\sc ii}]\,6717\AA \, emission from constant density model obtained for N4
4\,C are shown for comparison. The best-fit model line ratios for this model are given in Table  \ref{linesratio_N44C}.}
    \label{lineprof_mod_N44D}
\end{figure*}

In constant pressure model, the total pressure is kept constant throughout the cloud. The sum of gas pressure, line radiation pressure, turbulent pressure and the outward pressure of star light remain constant. At any particular region on the cloud, the resulting forces are due to the various contributions to the pressure balance. Therefore the cloud remains in hydrostatic equilibrium. 
The hydrostatic equilibrium model would balance the pressure gradient due to the kinetic energy and momentum carried by stellar photon flux, with the thermal gas pressure exerted by ionized gas. However, this model significantly changes the gas density and width of the ionization front. The constant density represents the total hydrogen density constant throughout the nebula, but electron and molecular fractions vary with depth. In Fig.\,\ref{lineprof_mod_N44D}, we compare the spatial profiles of various emission lines, [O\,{\sc i}]\,6300\AA, [O\,{\sc ii}]\,7318\AA, [O\,{\sc iii}]\,5007\AA, [N\,{\sc ii}]\,6584\AA, [S\,{\sc ii}]\,6717\AA \,obtained for N44\,D1 with constant density as well as constant pressure models. We note that constant pressure models provide a relatively narrower ionization fronts than constant density models. \citet{pellegrini07,pellegrini09} applied constant pressure models to Orion bar and M17 PDRs for a self-consistent simulation of H$^+$, H$^0$ and H$_2$ regions, including additional turbulent pressure and magnetic field. In that model, the physical depth, the separation of H$^0$, and H$_2$, and overall geometry depend on the gas density. In our constant pressure model, we include additional turbulence of 10\,km\,s$^{-1}$. This provides turbulent line broadening in addition to the thermal line broadening, and slightly increases the widths of the spatial profiles in the ionization front. \citet{pellegrini07,pellegrini09} found that the presence of a magnetic field significantly increases the physical width of the PDR, as a result of reduced gas density and increased photon path length. We calculate the constant pressure models consisting of gas pressure, and turbulent pressure.  We did not include the pressure due to the magnetic field, as currently there is no observational evidence of a magnetic field in N44. In Table\,\ref{linesratio_N44D1} we show the diagnostic line ratios obtained for both constant pressure and constant density model along with the observed line ratios of N44\,D1. We note that the constant pressure model describes most of the line ratios, however, we obtain reduced $\chi^2$ for most of the line ratios in the constant density model. We also note that geometry of O\,{\sc iii}]\,5007\AA\, spatial profile matches reasonably well with observation in the constant density model. The constant pressure model predicts a relatively larger electron density than observed. Our these studies show that the depth of the H$^+$ region, the width of ionization front, and overall geometry of the emission line profiles with depth are largely dependent on the thermal gas pressure and stellar radiation pressure. Within H$^+$ regions, the density can be constant with the depth; hence the predicted line ratios have a minimal effect on the chosen equation of state.

\subsection{Photoionization model of N44\,D1}
Even though observations give a photon escape fraction of 0.36 for N44\,D1, it does not mean that the H\,{\sc ii} region is completely ionization-bounded. We expect the photon leakage from some part of the cloud; hence that direction can be density-bounded, and the remaining part of the cloud can be ionization-bounded. This model can be a Blister type H\,{\sc ii} region as suggested by \citet{pellegrini11}. 
It is also interesting to note that N44\,D1 shows relatively large [O\,{\sc iii}]/H$\beta$ ratios, implying a large ionization parameter which can be a result of the high effective temperature of the only ionizing, O5\,V, star in N44\,D1. However, to obtain a high [O\,{\sc iii}]/H$\beta$ ratio, we include an additional contribution from a Bremsstrahlung component with the temperature in a range 10$^{6.2}$--10$^{6.4}$\,K. This plasma temperature is inferred from the studies of the X-ray emission by \citep{chu1993}. These authors have reported an excess of X-ray emission in the N44\,D region. This excess X-ray emission can be shock-driven due to massive stellar winds or the off-center supernovae remnant. We find that partial ionization-bounded geometry with constant density models can reproduce reasonably well the observed geometry and line ratios of N44\,D1 for a log\,$\Phi$(H)$=10.14$\,ph\,s$^{-1}$\,cm$^{-2}$, electron density 136\,cm$^{-3}$, and Bremsstrahlung temperature 1.67$\times 10^{6}$\,K. 
\subsection{Photoionization model of N44\,C}
The observed geometry of N44\,C does not give a well-defined ionization front at the H\,{\sc ii} region boundary, and the patches of [S\,{\sc ii}] and [O\,{\sc ii}] concentrations are found closer to the cloud centroid. We obtain an optimal model for N44\,C by applying an optically-thin constant density distribution. We note that the shape of the incident radiation mainly depends on three fixed stellar atmosphere models, which are appropriate for three enclosed hot stars O5\,III, O8.5\,V, and O9.5\,V. The [O\,{\sc iii}]/H$\beta$ ratio in N44\,C is relatively low compared to N44\,D1, indicating low ionization parameter and to obtain this gas ionization we did not include any additional Bremsstrahlung component as in N44\,D1. Since the observations show that 70$\%$ of photons escape from N44\,C, we use an optically thin open geometry where the Lyman continuum optical depth ($\tau_{912}$) is found to be very low ($<$\,1), hence the majority of the cloud is open to the Lyman photons. We note, this model can reasonably well describe the observed geometry and the line ratios of N44\,C. Fig.\,\ref{lineprof_mod_N44D} shows the model emission line profiles as a function of depth in the nebula, and the line ratios are given in Table\,\ref{linesratio_N44C}. Comparison of our models with the observed geometry and line ratios indicate that N44\,C has an optically thin geometry and the region is being energized mainly by the outward momentum carried by the radiation pressure from three ionizing stars.

\section{Conclusions}

We carry out a detailed analysis of two H\,{\sc ii} regions in N44 using the integral field optical spectroscopic observations obtained with MUSE. Comparing these observations with the photoionization models computed with CLOUDY, we describe the spatial distribution of emission line geometry and the physical conditions. 
Our results are summarized as follow:
\begin{enumerate}
\item Our analysis reveals that the spatial distribution of various spectral lines in N44\,D1 provides a stratified ionization geometry. The central ionizing star is covered by a fully ionized hydrogen gas and at the periphery, there is a well-defined transition zone from O$^{++}$ through O$^+$ to the neutral zone O$^0$. H$\alpha$, H$\beta$ and [O\,{\sc iii}] emission are co-spatial and peak at the fully ionized zone, while [O\,{\sc ii}], [N\,{\sc ii}], [S\,{\sc ii}] and [O\,{\sc i}] emission peak at the outer boundary. This region provides an excellent site for modeling an ideal H\,{\sc ii} region with a stratified ionization geometry. The [O\,{\sc i}]\,6300\AA\, emission of the N44 D1 region reveals a clear boundary/transition zone in the outer boundary of [O\,{\sc iii}]\,5007\AA\, and [O\,{\sc ii}]\,7329\AA\, emission, which does not cover the entire nebula completely, indicating a partial ionization front. Comparing these studies with \citet{pellegrini11, pellegrini12}, we suggest that the N44\,D1 is a Blister H\,{\sc ii} region.

\item The spatial distributions of various spectral lines in N44\,C do not show a stratified ionization front at the boundary. However, it shows the condensations of [S\,{\sc ii}] and [N\,{\sc ii}] emission within the H\,{\sc ii} region. The [O\,{\sc i}]\,6300\AA\, emission is relatively weak in N44\,C and does not show a well-defined outer boundary as in N44\,D1. These observations support relatively higher photon escape fraction reported by \citet{McLeod19}, suggesting that N44\,C is a density-bounded optically thin H\,{\sc ii} region. 

\item Our studies reveal that the [O\,{\sc iii}]/H$\alpha$ and [O\,{\sc iii}]/H$\beta$ line ratios give a clear indication of a higher degree of ionization in the regions closer to the ionizing stars and the ratios are lower toward the boundary of N44\,D1 and N44\,C. [S\,{\sc ii}]/H$\alpha$ and [N\,{\sc ii}]/H$\beta$ line ratio maps show a shell structure in both N44\,D1 and N44\,C. [O\,{\sc iii}]/H$\alpha$ and [O\,{\sc iii}]/H$\beta$ of N44\,D1 are much higher than N44\,C, indicating a harder radiation field. The effective temperature of the hot star plays a key role here because N44\,D1 has a hotter ionizing star (O5\,V) than N44\,C (O5\,III).

\item We use our results of spatially resolved MUSE spectra to explore the photoionization models with CLOUDY that can well describe the observed geometry and emission line ratios. We find that the constant density model gives better geometry and line ratios than the constant pressure model in the N44\,D1. An ionization-bounded geometry with a partial covering factor can well reproduce the observed geometry and line ratios, indicating that N44\,D1 is a Blister H\,{\sc ii} region. The spatial profile of [O\,{\sc iii}]\,5007\AA\, matches very nicely with the observation. Model calculations reveal that a significant amount of X-ray emission takes part in shaping the geometry of the emission line profiles, in addition to the ionizing radiation from the O5\,V star. The electron density and temperature values from this model are also consistent with our measurements from the observed values. 

\item An optically thin and open geometry model has been applied to reproduce the observed geometry and line ratios in N44\,C. The modeling results show that N44\,C region is mainly energized by the radiation from three ionizing stars. Our studies indicate that the ionization structure and physical conditions in N44\,D1 and N44\,C are set by the stellar radiation pressure and gas thermal pressure.

\end{enumerate}
\section*{Acknowledgments}

This research has been supported by the United Arab Emirates University (UAEU) through UAEU Program for Advanced Research (UPAR) grant G00003479 and start-up grant G00002964. This paper makes use of the following MUSE (VLT) data: program ID: 096.C--0137(A). F. Kemper acknowledges the Ministry of Science and Technology of Taiwan for the grant MOST107-2119-M-001-031-MY3 and Academia Sinica Investigator Award, AS-IA-106-M03. M. Sewilo acknowledges the NASA award, 80GSFC21M0002 (M. S.).
\begin{table*}
\centering
\caption{Luminosities of observed emission lines }
\renewcommand{\arraystretch}{1.2}
\begin{tabular}{lcccccccc}
\hline
\multicolumn{1}{c}{Emission lines}&
\multicolumn{2}{c}{N44\,D1}&
\multicolumn{2}{c}{N44\,D2}&
\multicolumn{2}{c}{N44\,C}&\\
&$L_{\textrm{obs}}$(erg s$^{-1}$)$^{\text{a}}$&$L_{\textrm{int}}$(erg s$^{-1}$)$^{\text{b}}$&$L_{\textrm{obs}}$(erg s$^{-1}$)&$L_{\textrm{int}}$(erg s$^{-1}$)&$L_{\textrm{obs}}$(erg s$^{-1}$)&$L_{\textrm{int}}$(erg s$^{-1}$)\\
&$\times$10$^{36}$&$\times$10$^{36}$&$\times$10$^{36}$&$\times$10$^{36}$&$\times$10$^{36}$&$\times$10$^{36}$&\\
\hline
H$\alpha$       &14.70\,$\pm$\,1.62 &17.5\,$\pm$\,1.93 &5.36\,$\pm$\,0.58 &5.45\,$\pm$\,0.69 &13.9\,$\pm$\,1.53 &21.5\,$\pm$\,1.86\\
H$\beta$                            &4.68\,$\pm$\,0.47 &6.12\,$\pm$\,0.61 &1.71\,$\pm$\,0.24 &1.76\,$\pm$\,0.31 &3.69\,$\pm$\,0.55 &7.24\,$\pm$\,0.75\\
\big[N {\sc ii}\big]6584\AA         &0.80\,$\pm$\,0.13 &0.96\,$\pm$\,0.16 &0.29\,$\pm$\,0.07 &0.30\,$\pm$\,0.08 &0.97\,$\pm$\,0.16 &1.50\,$\pm$\,0.20\\
\big[O {\sc i}\big]6300\AA          &0.32\,$\pm$\,0.07 &0.39\,$\pm$\,0.08 &0.07\,$\pm$\,0.03 &0.07\,$\pm$\,0.03 &0.13\,$\pm$\,0.04 &0.21\,$\pm$\,0.05\\
\big[O {\sc ii}\big]7318\AA         &0.16\,$\pm$\,0.05 &0.19\,$\pm$\,0.06 &0.06\,$\pm$\,0.03 &0.06\,$\pm$\,0.03 &0.17\,$\pm$\,0.05 &0.24\,$\pm$\,0.06\\
\big[O {\sc ii}\big]7329\AA         &0.15\,$\pm$\,0.04 &0.17\,$\pm$\,0.04 &0.05\,$\pm$\,0.02 &0.05\,$\pm$\,0.02 &0.14\,$\pm$\,0.04 &0.20\,$\pm$\,0.04\\
\big[O {\sc iii}\big]4959\AA        &12.40\,$\pm$\,0.64 &16.1\,$\pm$\,2.13 &2.98\,$\pm$\,0.38 &3.06\,$\pm$\,0.50 &2.61\,$\pm$\,0.04 &5.02\,$\pm$\,0.56\\
\big[O {\sc iii}\big]5007\AA        &36.90\,$\pm$\,3.85 &48.5\,$\pm$\,5.06 &8.87\,$\pm$\,0.99 &9.11\,$\pm$\,1.30 &7.66\,$\pm$\,1.09 &15.1\,$\pm$\,1.48\\
\big[S {\sc ii}\big]6717\AA         &1.11\,$\pm$\,0.16 &1.31\,$\pm$\,0.18 &0.34\,$\pm$\,0.07 &0.35\,$\pm$\,0.08 & 0.94\,$\pm$\,0.15&1.43\,$\pm$\,0.18\\
\big[S {\sc ii}\big]6732\AA         &0.83\,$\pm$\,0.13 &0.98\,$\pm$\,0.15 &0.26\,$\pm$\,0.06 &0.26\,$\pm$\,0.07 &0.68\,$\pm$\,0.11 &1.04\,$\pm$\,0.14\\
\big[S {\sc iii}\big]9069\AA        &1.48\,$\pm$\,0.21 &1.62\,$\pm$\,0.23 &0.60\,$\pm$\,0.10 &0.61\,$\pm$\,0.11 &1.45\,$\pm$\,0.18 &1.82\,$\pm$\,0.20\\

\\
\hline
\label{lines}
\end{tabular}
\parbox{160 mm}{a: $L_{\textrm{obs}}$ is the observed luminosity.\\ b: $L_{\textrm{int}}$ is the reddening corrected luminosity.}
\end{table*}

\begin{table*}
\centering
\caption{Emission line properties}
\setlength{\tabcolsep}{11pt}
\renewcommand{\arraystretch}{1.5}
\begin{tabular}{@{}lccccccc}
\hline
\multicolumn{1}{c}{Regions}&
\multicolumn{1}{c}{$Q$}&
\multicolumn{1}{c}{$Q_{\textrm{o}}$}&
\multicolumn{1}{c}{$f_{\textrm{esc}}$}&
\multicolumn{1}{c}{$\langle n_{\textrm{e}}\rangle ^{\text{a}}$}&
\multicolumn{1}{c}{$n_{\textrm{e}}$ [S {\sc ii}] $^{\text{b}}$}&
\multicolumn{1}{c}{$n_{\textrm{e}}$ [S {\sc ii}] $^{\text{c}}$}\\
&(s$^{-1}$)&(s$^{-1}$)&&(cm$^{-3}$)&(cm$^{-3}$)&(cm$^{-3}$)&\\
\hline
N44\,D1&1.07$\times 10^{49}$&1.66 $\times 10^{49}$&0.36&31&132$\pm$50& 141$\pm$43\\
N44\,D2&0.36$\times 10^{49}$&1.26$\times 10^{49}$&0.71&26&115$\pm$45& 121$\pm$37\\
N44\,C&1.02$\times 10^{49}$&3.37$\times 10^{49}$&0.70 &38 &66$\pm$40 & 92$\pm$35\\
\hline
\label{properties}
\end{tabular}
\parbox{150 mm}{a: $\langle n_{\textrm{e}}\rangle$ is the average electron density from H$\alpha$ emission.\\ b: $n_{\textrm{e}}$[S {\sc ii}] is the electron density derived using PYNEB.\\ c: $n_{\textrm{e}}$[S {\sc ii}] is the electron density derived from the equation no. 11.}
\end{table*}

\begin{table*}
\centering
\caption{Model line ratios compared with observations for N44\,D1}
\renewcommand{\arraystretch}{1.2}
\begin{tabular}{lcccccccccc}
\hline
\multicolumn{1}{c}{Line ratios}
&
\multicolumn{1}{c}{Observed}&
\multicolumn{1}{c}{Pressure Model}&
\multicolumn{1}{c}{$\chi^{2}$}&
\multicolumn{1}{c}{Density Model}&
\multicolumn{1}{c}{$\chi^{2}$}&\\
&ratios &$\Phi$(H)$_{10.15}$& & $\Phi$(H)$_{10.14}$& \\
\hline
\big[S {\sc ii}\big]\,6717/6732                        &1.32$\pm$0.40    &1.28& 0.01  &1.32&  0.00\\
\big[O {\sc ii}\big]\,(7318+7329)/[O {\sc iii}]5007     &0.008$\pm$0.002&0.004 & 4.0 &0.006&  1.00\\
\big[O {\sc iii}\big]\,4959/5007                       &0.33$\pm$0.07   &0.33& 0.00  &0.33&  0.00\\
\big[O {\sc iii}\big]\,5007/H$_\beta$                   &7.89$\pm$1.60     &6.76& 0.50  &7.10&  0.24\\
\big[S {\sc iii}\big]\,9069/H$_\beta$                  &0.31$\pm$0.08    &0.36& 0.40 &0.37&  0.56\\
\big[O {\sc ii}\big]\,7329/H$_\beta$                   &0.03$\pm$0.01    &0.02& 1.0 &0.02&  1.00\\
\big[S {\sc ii}\big]\,6717/H$_\beta$                   &0.24$\pm$0.10   &0.36& 1.44 &0.34&  1.00\\
\big[N {\sc ii}\big]\,6584/H$_\beta$                   &0.17$\pm$0.05   &0.14& 0.36 &0.14&  0.36\\
\big[O {\sc i}\big]\,6300/H$_\beta$                    &0.07$\pm$0.05   &0.15& 2.56 &0.14&  1.96\\
\big[O {\sc iii}\big]\,5007/H$_\alpha$                  &2.52$\pm$0.50    &2.40& 0.06  &2.51&  0.00\\
\big[S {\sc ii}\big]\,6717/H$_\alpha$                  &0.08$\pm$0.02   &0.13& 6.30 &0.12&  4.00\\
\big[N {\sc ii}\big]\,6584/H$_\alpha$                  &0.05$\pm$0.02   &0.05& 0.00 &0.05&  0.00\\
$T_{\textrm{e}}$ over radius (K)                            & &12\,491&  &12\,410& \\
$n_{\textrm{e}}$ over radius (cm$^{-3}$)                    & &175&    &136& 
 \\
\hline
\label{linesratio_N44D1}
\end{tabular}
\vspace{-8mm}
\end{table*}

\begin{table*}
\centering
\caption{Model line ratios compared with observations for N44\,C}
\renewcommand{\arraystretch}{1.2}
\begin{tabular}{lcccc}
\hline
\multicolumn{1}{c}{Line ratios}&
\multicolumn{1}{c}{Observed ratios}&
\multicolumn{1}{c}{Model ratios}&
\multicolumn{1}{c}{$\chi^{2}$}&\\
\hline
\big[S {\sc ii}\big]\,6717/6732               &1.37$\pm$0.4    &1.35 &  0.002 \\
\big[O {\sc ii}\big]\,(7318+7329)/[O {\sc iii}]5007   &0.04$\pm$0.02   &0.02 &  1.00\\
\big[O {\sc iii}\big]\,4959/5007              &0.34$\pm$0.01   &0.33 &   1.00\\
\big[O {\sc iii}\big]\,5007/H$_\beta$          &2.07$\pm$0.6    &1.75&   0.28 \\
\big[S {\sc iii}\big]\,9069/H$_\beta$         &0.39$\pm$0.1   & 0.28 &   1.20\\
\big[O {\sc ii}\big]\,7329/H$_\beta$          &0.03$\pm$0.02   &0.02 &   0.25\\
\big[S {\sc ii}\big]\,6717/H$_\beta$          &0.25$\pm$0.08   &0.30 &  0.39\\
\big[N {\sc ii}\big]\,6584/H$_\beta$          &0.26$\pm$0.08   &0.19 &  0.76\\
\big[O {\sc i}\big]\,6300/H$_\beta$           &0.03$\pm$0.01   &0.03 &  0.00\\
\big[O {\sc iii}\big]\,5007/H$_\alpha$         &0.55$\pm$0.14  &0.63 &  0.32 \\
\big[S {\sc ii}\big]\,6717/H$_\alpha$         &0.07$\pm$0.02   &0.10 &  2.25\\
\big[N {\sc ii}\big]\,6584/H$_\alpha$         &0.07$\pm$0.02   &0.07 &   0.00 \\
$T_{\textrm{e}}$ over radius (K)      &&10\,709 & \\
$n_{\textrm{e}}$ over radius (cm$^{-3}$)                     &&84 & \\
\hline
\label{linesratio_N44C}
\end{tabular}
\vspace{-8mm}
\end{table*}

\bibliography{n44_draft_v1}{}
\bibliographystyle{aasjournal}


\end{document}